\documentclass[11pt]{amsart}
\usepackage{amsbsy,amssymb,amscd,amsfonts,latexsym,amstext,delarray,
amsmath,graphicx, epsfig}
 \usepackage[ansinew]{inputenc}
\usepackage{color}
\input xypic

\def\taille{\long\def\epsfsize##1##2{\facteur\textwidth}}
\def\fig#1#2{\long\def\facteur{#1}\taille\epsffile{#2}}

\newtheorem{thm}{Theorem}[section]

\newtheorem{conj}[thm]{Conjecture}

\def\bF{{\mathbb F}}
\def\bG{{\mathbb G}}

\def\bL{{\mathbb L}}

\def\bQ{{\mathbb Q}}

\def\bT{{\mathbb T}}
\def\bU{{\mathbb U}}

\def\A{{\mathbb A}}
\def\C{{\mathbb C}}

\def\N{{\mathbb N}}
\renewcommand{\P}{{\mathbb P}}
\def\Q{{\mathbb Q}}
\def\Z{{\mathbb Z}}
\def\R{{\mathbb R}}
\def\K{{\mathbb K}}

\def\cancel#1#2{\ooalign{$\hfil#1\mkern1mu/\hfil$\crcr$#1#2$}}
\def\Dirac{\mathpalette\cancel D}

\def\m{{\mathfrak m}}

\def\cA{{\mathcal A}}

\def\cC{{\mathcal C}}
\def\cD{{\mathcal D}}
\def\cE{{\mathcal E}}
\def\cF{{\mathcal F}}

\def\cH{{\mathcal H}}

\def\cK{{\mathcal K}}
\def\cL{{\mathcal L}}
\def\cM{{\mathcal M}}
\def\cN{{\mathcal N}}

\def\cP{{\mathcal P}}

\def\cR{{\mathcal R}}
\def\cS{{\mathcal S}}
\def\cT{{\mathcal T}}

\def\cV{{\mathcal V}}

\newcommand{\ie}{{\it i.e.\/}\ }
\newcommand{\eg}{{\it e.g.\/}\ }
\newcommand{\cf}{{\it cf.\/}\ }

\def\text{\hbox}

\def\Aut{{\rm Aut}}

\def\End{{\rm End}}
\def\Ext{{\rm Ext}}

\def\GL{{\rm GL}}
 
\def\Hom{{\rm Hom}}

\def\Lie{{\rm Lie}}

\def\Spec{{\rm Spec}}

\def\Tr{{\rm Tr}}

\title{Feynman integrals and motives}
\author{Matilde Marcolli}
\date{}

\begin{document}

\maketitle

\begin{verse}
{\em and Beyond the Infinite} \\
(Stanley Kubrick, {\em 2001 - A Space Odyssey})
\end{verse}

\begin{abstract}
This article gives an overview of recent results on the relation between
quantum field theory and motives, with an emphasis on two different
approaches: a ``bottom-up" approach based on the algebraic geometry
of varieties associated to Feynman graphs, and a ``top-down" approach
based on the comparison of the properties of associated
categorical structures. This survey is mostly based on joint work
of the author with Paolo Aluffi, along the lines of the first approach,
and on previous work of the author with Alain Connes on the second
approach.  
\end{abstract}

\tableofcontents

\section{Introduction: quantum fields and motives, an unlikely match}

This paper, based on the plenary lecture delivered by the author at the
5th European Congress of Mathematics in Amsterdam, aims at giving an
overview of the current approaches to understanding the role of motives
and periods of motives in perturbative quantum field theory. 
It is a priori surprising that there should be any relation at all
between such distant fields. In fact, motives are a very abstract
and sophisticated branch of algebraic and arithmetic geometry,
introduced by Grothendieck as a universal cohomology theory
for algebraic varieties. On the other hand, perturbative quantum
field theory is a procedure for computing, by successive 
approximations in powers of the relevant coupling constants, 
values of physical observables in a quantum field theory. 
Perturbative quantum field theory is not entirely mathematically 
rigorous, though as we will see later in this paper, a lot of interesting 
mathematical structures arise when one tries to understand 
conceptually the procedure of extraction of finite values from
divergent Feynman integrals known as renormalization.

The theory of motives itself has its mysteries, which make it
a very active area of research in contemporary mathematics.
The categorical structure of motives is still a problem very much 
under investigation. While one has a good abelian category of 
pure motives (with numerical equivalence), that is, of motives
arising from smooth projective varieties, the ``standard conjectures" 
of Grothendieck are still unsolved. Moreover, when it comes to the much 
more complicated setting of mixed motives, which no longer correspond 
to smooth projective varieties, one knows that they form a triangulated 
category, but in general one cannot improve that to the
level of an abelian category with the same nice properties one has
in the case of pure motives. See \cite{BloMM}, \cite{Levine} for an 
overview of the theory of mixed motives.

The unlikely interplay between motives and quantum field theory
has recently become an area of growing interest at the interface of
algebraic geometry, number theory, and theoretical physics. 
The first substantial indications of a relation between these
two subjects came from extensive
computations of Feynman diagrams carried out by Broadhurst
and Kreimer \cite{BroKr}, which showed the presence of
multiple zeta values as results of Feynman integral calculations. 
{}From the number theoretic viewpoint, multiple zeta values are a
prototype case of those very interesting classes of numbers which, 
altough not themselves algebraic, can be
realized by integrating algebraic differential forms on 
algebraic cycles in arithmetic varieties. Such numbers are
called periods, \cf \cite{KoZa}, and there are precise conjectures
on the kind of operations (changes of variables, Stokes formula)
one can perform at the level of the algebraic data that
will correspond to relations in the algebra of periods. As one 
can consider periods of algebraic varieties, one can also consider
periods of motives. In fact, the nature of the numbers one obtains
is very much related to the motivic complexity of the part of the
cohomology of the variety that is involved in the
evaluation of the period.

There is a special class of motives that are better understood and
better behaved with respect to their categorical properties: the {\em mixed 
Tate motives}. They are also the kind of motives that are expected (see
\cite{Gon2}, \cite{Tera}) to be supporting the type of periods like 
multiple zeta values that appear in Feynman integral computations. 

At the level of pure motives the Tate motives
$\Q(n)$ are simply motives of projective spaces and their formal inverses,
but in the mixed case there are very nontrivial extensions of
these objects possible. In terms of algebraic varieties, for instance,
varieties that have stratifications where the successive strata are obtained
by adding copies of affine spaces provide examples of mixed Tate motives. 
There are various conjectural geometric descriptions of such
extensions (see \eg \cite{BGSV} for one possible description
in terms of hyperplane arrangements). 
Understanding when certain geometric objects determine motives that
are or are not mixed Tate is in general a difficult question and, it turns out,
one that is very much central to the relation to quantum field theory. 

In fact, the main conjecture we describe here, along with an overview of
some of the current approaches being developed to answer it, is whether,
after a suitable subtraction of infinities, the Feynman integrals of a 
perturbative scalar quantum field theory always produce values that are
periods of mixed Tate motives. 

\subsection{Feynman diagrams: graphs and integrals}

We briefly introduce the main characters of our story, starting with Feynman
diagrams. By these one usually means the data of a finite graph together with 
a prescription for assigning variables to the edges with linear relations at the 
vertices and a formal integral in the resulting number of independent variables. 

For instance, consider a graph of the form

\centerline{\fig{0.5}{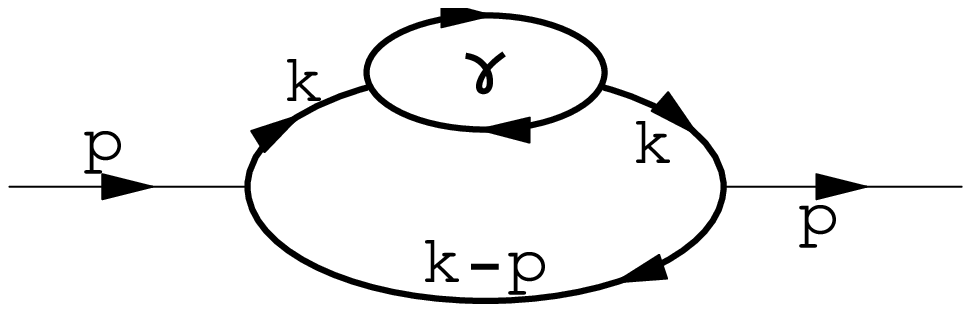}}

The corresponding integral gives
$$ (2\pi)^{-2D}\int \frac{1}{k^4}\,\frac{1}{(k-p)^2}\,\frac{1}{(k+\ell)^2}\,
\frac{1}{\ell^2}\,d^Dk\,d^D\ell $$
As is often the case, the resulting integral is divergent. We will explain below the regularization procedure that expresses such divergent integrals in terms of meromorphic functions. In this
case one obtains
$$ (4\pi)^{-D}\,\frac{\Gamma(2-\frac{D}{2})\Gamma(\frac{D}{2}-1)^3\Gamma(5-D)
\Gamma(D-4)}
{\Gamma(D-2)\Gamma(4-\frac{D}{2})\Gamma(\frac{3D}{2}-5)}\,(p^2)^{D-5} $$
and one identifies the divergences with poles of the resulting function.

The renormalization problem in perturbative quantum field theory consists of
removing the divergent part of such expressions by a redefinition of the running
parameters (masses, coupling constants) in the Lagrangian of the theory.
To avoid non-local expressions in the divergences, which cannot be canceled 
using the local terms in the Lagrangian, one needs a method to remove divergences
from Feynman integrals that accounts for the nested structure of subdivergences
inside a given Feynman graphs. Thus, the process of extracting finite
values from divergent Feynman integrals is organized in two steps:
{\em regularization}, by which one denotes a procedure that replaces a divergent
integral by a function of some new regularization parameters, which is meromorphic
in these parameters, and happens to have a pole at the value of the parameters
that recovers the original expression; and {\em renormalization}, which denotes 
the procedure by which the polar part of the Laurent series obtained as a 
result of the regularization process is extracted {\em consistently} with the 
hierarchy of divergent subgraphs inside larger graphs.

\subsection{Perturbative Quantum Field Theory in a nutshell}

We recall very briefly here a few notions of perturbative quantum field
theory we need in the following. A detailed introduction for the use of
mathematicians is given in Chapter 1 of \cite{CoMa-book}.

To specify a quantum field theory, which we denote by 
$\cT$ in the following, one needs to assign the Lagrangian of the 
theory. We restrict ourselves to the case of {\em scalar theories},
though it is possible that similar conjectures on number theoretic
aspects of values of Feynman integrals may be formulated more
generally.

A scalar field theory $\cT$ in spacetime dimension $D$ is determined by
a classical Lagrangian density of the form
\begin{equation}\label{Lagrangian}
 \cL(\phi)=\frac{1}{2} (\partial \phi)^2 + \frac{m^2}{2} \phi^2 
+ \cL_{int}(\phi), 
\end{equation}
in a single scalar field $\phi$, with the interaction term
$\cL_{int}(\phi)$ given by a polynomial in $\phi$ of degree at least three.
This determines the corresponding classical action as
$$ S(\phi)=\int \cL(\phi) d^Dx =S_0(\phi)+S_{int}(\phi). $$

While the variational problem for the classical action gives the
classical field equations, the quantum corrections are implemented by
passing to the {\em effective action} $S_{eff}(\phi)$. The latter is not
given in closed form, but in the form of an asymptotic series, the
perturbative expansion parameterized by the ``one-particle irreducible" (1PI)
Feynman graphs. The resulting expression for the effective action is then of
the form
\begin{equation}\label{Seff}
 S_{eff}(\phi)=S_0(\phi)+\sum_\Gamma \frac{\Gamma(\phi)}{\#\Aut(\Gamma)} 
\end{equation} 
where the contribution of a single graph is an integral on external momenta
assigned to the ``external edges" of the graph,
$$ \Gamma(\phi)=\frac{1}{N!} \int_{\sum_i p_i=0} \hat\phi(p_1)\cdots
\hat\phi(p_N) U^z_\mu(\Gamma(p_1,\ldots,p_N)) dp_1\cdots dp_N. $$
In turn, the function of the external momenta that one integrates to obtain
the coefficient $\Gamma(\phi)$ is an integral in momentum variables
assigned to the ``internal edges" of the graph $\Gamma$, with momentum
conservation at each vertex. Thus, it can be expressed as an integral in a
number of variables equal to the number $b_1(\Gamma)$ of loops in the graph,
of the form
\begin{equation}\label{UGammaP}
 U(\Gamma(p_1,\ldots,p_N))=\int I_\Gamma(k_1,\ldots,k_\ell,p_1,\ldots,p_N) 
d^Dk_1 \cdots d^D k_\ell . 
\end{equation}
The graphs involved in the expansion \eqref{Seff} are the 1PI Feynman graphs
of the theory $\cT$, \ie those graphs that cannot be disconnected by the
removal of a single edge. As Feynman graphs of a given theory, they are also
subject to certain combinatorial constraints: each
vertex in the graph has valence equal to the degree of one of the monomials
in the Lagrangian. The edges are subdivided into internal edges connecting
two vertices and external edges (or half edges) connected to a single vertex.
The Feynman rules of the theory $\cT$ specify how to assign an integral \eqref{UGammaP}
to a Feynman graph, namely it specifies the form of the function
$I_\Gamma(k_1,\ldots,k_\ell,p_1,\ldots,p_N)$ of the internal momenta. This is
a product of ``propagators" associated to the internal lines. These are typically of the form
$1/q(k)$, where $q$ is a quadratic form in the momentum variable of a given internal edge,
which is obtained from the fundamental (distributional) solution of the associated classical 
field equation for the free field theory coming from the $S_0(\phi)$ part of the Lagrangian, 
such as the Klein--Gordon equations for the scalar case. Momentum conservations
are then imposed at each vertex, and multiplied by a power of the coupling constant 
(the coefficient of the corresponding monomial in the Lagrangian) and a power of $2\pi$.

As we mentioned above, the resulting integrals \eqref{UGammaP} are very often divergent.
Thus, a regularization and renormalization method is used to extract a finite value. There are
different regularization and renormalization schemes used in the physics literature. We
concentrate here on Dimensional Regularization and Minimal Subtraction, which is
a widely used regularization method in particle physics computations, and on the recursive
procedure of Bogolyubov--Parasiuk--Hepp--Zimmermann for renormalization
\cite{BoPa}, \cite{Hepp}, \cite{Zimm}, see also \cite{Manou}. Regularization
and Renormalization are two distinct steps in the process of extracting finite values from
divergent Feynman integrals. The first replaces the integrals with meromorphic functions
with poles that account for the divergences, while the latter organizes subdivergences in
such a way that the divergent parts can be eliminated (in the case of a renormalizable
theory) by readjusting finitely many parameters in the Lagrangian.

The procedure of Dimensional Regularization is based on the curious idea of
making sense of the integrals \eqref{UGammaP} in ``complexified dimension"
$D-z$, with $z\in \C^*$, instead of working in the original dimension $D\in \N$.
It would seem at first that, to make sense of such a procedure, one would need
to make sense of geometric spaces in dimension $D-z$ and of a corresponding
theory of measure and integration in such spaces. However, due to the special
form of the Feynman integrals \eqref{UGammaP}, a lot less is needed. In fact,
it turns out that it suffices to have a {\em formal procedure} to define the Gaussian
integral 
\begin{equation}\label{Gaussian}
\int e^{-\lambda t^2} d^D t := \pi^{D/2} \lambda^{-D/2}
\end{equation}
in the case where $D$ is no longer a positive integer but a complex number.
Clearly, since the right hand side of \eqref{Gaussian} continues to make sense
for $D\in \C^*$, one can use that as the definition of the left hand side and set:
\begin{equation}\label{zGaussian}
\int e^{-\lambda t^2} d^z t := \pi^{z/2} \lambda^{-z/2},  \ \ \ \  \forall z\in \C^*.
\end{equation}
The computations of Feynman integrals can be reformulated in terms
of Gaussian integrations using the method of Schwinger parameters we
return to in more detail below, hence one obtains a well defined notion of
integrals in dimension $D-z$:
\begin{equation}\label{zUGammaP}
U^z_\mu(\Gamma(p_1,\ldots,p_N))  = \int \mu^{z\ell} 
d^{D-z} k_1 \cdots d^{D-z} k_\ell  I_\Gamma(k_1,\ldots,k_\ell,p_1,\ldots,p_N) .
\end{equation}
The variable $\mu$ has the physical units of a mass and appears
in these integrals for dimensional reasons. It will play an important role later on,
as it sets the dependence on the energy scale of the renormalized values of
the Feynman integrals, hence the renormalization group flow.

It is not an easy result to show that the dimensionally regularized integrals
give meromorphic functions in the variable $z$, with a Laurent series
expansion at $z=0$. See a detailed discussion of this point in Chapter 1 of
\cite{CoMa-book}. We will not enter in details here and talk loosely about
\eqref{zUGammaP} as a meromorphic function of $z$ depending on the
additional parameter $\mu$.

We return to a discussion of a possible geometric meaning of the dimensional
regularization procedure in the last section of this paper.

\subsection{The Feynman rules}

The integrand $I_\Gamma(k_1,\ldots,k_\ell,p_1,\ldots,p_N)$ in the Feynman
integrals \eqref{UGammaP} is determined by the {\em Feynman rules} of the
given quantum field theory, see \cite{ItZu}, \cite{BjDr}.
These can be summarized as follows:
\begin{itemize}
\item A Feynman graph $\Gamma$ of a scalar quantum field theory with
Lagrangian \eqref{Lagrangian} has vertices of valences equal to the degrees
of the monomials in the Lagrangian, internal edges connecting pairs of vertices,
and external edges connecting to a single vertex.
\item To each internal edge of a Feynman graph $\Gamma$ one assigns a
momentum variable $k_e\in \R^D$ and a propagator, which is a quadratic form 
$q_e$ in the variable $k_e$, which (in Euclidean signature) is of the form
\begin{equation}\label{propagator}
 q_e(k_e)= k_e^2 + m^2 .
\end{equation} 
\item The integrand is obtained by taking a product over all internal edges 
of the inverse propagators
$$ \frac{1}{q_1 \cdots q_n}  $$
and imposing a linear relation at each vertex, which expresses the conservation
law 
$$ \sum_{e_i \in E(\Gamma):s(e_i)=v} k_i =0 $$
for the momenta flowing through that vertex. One obtains in this way the
integrand
\begin{equation}\label{FeyruleIGamma}
 I_\Gamma(k_1,\ldots,k_\ell,p_1,\ldots,p_N) =
\frac{\delta(\sum_{i \in E_{int}(\Gamma)} \epsilon_{v,i} k_i + \sum_{j\in E_{ext}(\Gamma)}
 \epsilon_{v,j} p_j)}{ q_1(k_1) \cdots q_n(k_n) } ,
\end{equation}
where $\epsilon_{e,v}$ denotes the incidence matrix of the graph
$$ \epsilon_{e,v}=\left\{ \begin{array}{rl} +1 & t(e)=v \\
-1 & s(e)=v \\
0 & \text{otherwise.} \end{array} \right. $$ 
\item For each vertex of $\Gamma$ one also multiplies the above by a constant factor 
involving the coupling constants of the terms in the Lagrangian of power corresponding
to the valence of the vertex and by a power of $(2\pi)$, which we omit for simplicity.
\end{itemize}

There are two properties of Feynman rules that it is useful to recall for 
comparison with algebro-geometric settings:
\begin{enumerate}
\item Reduction from graphs to connected graphs: the Feynman rules are
multiplicative over disjoint unions of graphs
\begin{equation}\label{Feyrules1}
U(\Gamma,p) =U(\Gamma_1,p_1) \,\, U (\Gamma_2, p_2) , \ \ \ \text{ for }
\ \ \ \Gamma = \Gamma_1 \amalg \Gamma_2.
\end{equation}
\item Reduction from connected graphs to 1PI graphs. An
arbitrary connected finite graph can be written as a tree $T$ where some of
the vertices are replaced by 1PI graphs with a number of external edges matching
the valence of the vertex, $\Gamma = \cup_{v \in V(T)} \Gamma_v$. For these
graphs the Feynman rules satisfy
\begin{equation}\label{Feyrules2}
U(\Gamma) = \prod_{v\in V(T)} U(\Gamma_v) \prod_{e\in E_{ext}(\Gamma_v),
e'\in E_{ext}(\Gamma_{v'}), e = e' \in E_{int}(\Gamma)}  \frac{\delta(p_e-p_{e'})}{q_{e}(p_e)} 
\end{equation}
\end{enumerate}
These properties reduce the combinatorics of Feynman graphs to the 1PI case.
Notice that in the particular case where $m\neq 0$ (massive theories)
and the external momenta are set to zero,  $p=0$, the case \eqref{Feyrules2}
reduces to the simpler form
\begin{equation}\label{Feyrules2b}
U(\Gamma) = U(L)^{\# E(T)} \prod_{v\in V(T)} U(\Gamma_v) , 
\end{equation}
where $U(L)$ is the inverse propagator for a single edge, in this case
just equal to the constant factor $m^{-2}$.

\subsection{Parametric representation of Feynman integrals}

The Feynman parameterization (also known as $\alpha$-parameterization), see \cite{BjDr}, \cite{ItZu}, \cite{Naka}, reformulates the Feynman integrals \eqref{UGammaP} 
in such a way that they become manifestly (modulo divergences)
written as the integral of an algebraic differential form on an algebraic variety,
integrated over a cycle with boundary on a divisor in the variety, see \cite{BEK}.

One starts with the Feynman integral, written as above in the form
$$ U(\Gamma)=\int \frac{\delta(\sum_{i=1}^n \epsilon_{v,i} k_i +\sum_{j=1}^N \epsilon_{v,j} p_j)} {q_1\cdots q_n} \,\, d^D k_1\cdots d^D k_n  $$
with $n=\# E_{int}(\Gamma)$ and $N=\# E_{ext}(\Gamma)$ and with $\epsilon_{e,v}$
the incidence matrix.

Then, one introduces the {\em Schwinger parameters}. These are variables $s_i \in \R_+$ 
defined by the identity
$$ q_1^{-k_1}\cdots q_n^{-k_n} =
 \frac{1}{\Gamma(k_1) \cdots \Gamma(k_n)} \int_0^\infty \cdots \int_0^\infty e^{-(s_1 q_1 +\cdots + s_n q_n)}\, s_1^{k_1-1} \cdots s_n ^{k_n-1}
 \,\,ds_1 \cdots ds_n. $$ 
The {\em Feynman trick}, which consists of writing
$$ \frac{1}{q_1 \cdots q_n} =(n-1)! \, 
\int \frac{\delta(1-\sum_{i=1}^n t_i)}{(t_1 q_1 + \cdots +
t_n q_n)^n} \,\, dt_1\cdots dt_n, $$ 
is obtained from a particular case of the identity defining the
Schwinger parameters, after a simple change of variables.

One then further introduces a change of variables $k_i = u_i + \sum_{k=1}^\ell \eta_{ik} x_k$, where
$ \eta_{ik}$ is the matrix
$$ \eta_{ik}=\left\{ \begin{array}{rl} +1 & \text{edge $e_i\in$ loop
$l_k$, same orientation} \\[2mm] -1 & \text{edge $e_i\in$ loop
$l_k$, reverse orientation} \\[2mm] 0 & \text{otherwise.} \end{array}\right.$$
This depends on the choice of an orientation of the edges and of a basis
of loops, \ie a basis of $H_1(\Gamma)$. The equations imposing the conservation
laws for momenta at each vertex, together with the constraint $\sum_i t_i u_i \eta_{ir} =0$
determine uniquely $u_i$ as functions of the external momenta $p$ and give
$$ \sum_i t_i u_i^2 = p^\dag R_\Gamma(t) p , $$
where $R_\Gamma(t)$ is a function defined in terms of the combinatorics of the
graph. Thus, one rewrites the Feynman integral after this change of coordinates
in the form
\begin{equation}\label{UGammaParamFey}
U(\Gamma)=   \frac{\Gamma(n-D\ell/2)}{(4\pi)^{\ell D/2}}  
\int_{\sigma_n} \frac{\omega_n}
{\Psi_\Gamma(t)^{D/2}V_\Gamma(t,p)^{n- D\ell/2}},  
\end{equation}
where $\omega_n$ is the volume form and the domain of integration is 
the simplex $\sigma_n=\{ t\in \R^n_+ | \sum_i t_i=1\}$.
In the {\em massless case} (with $m=0$) the term  $V_\Gamma(t,p) =p^\dag R_\Gamma(t) p + m^2$ 
is of the form
$$ V_\Gamma(t,p)|_{m=0} = \frac{P_\Gamma(t,p)}{\Psi_\Gamma(t)}, $$
where $P_\Gamma(t,p)$ is a homogeneous polynomial of degree $b_1(\Gamma)+1$ in $t$,
defined in terms of the cut-sets of the graph (complements of spanning tree plus one edge),
$$ P_\Gamma(t,p) = \sum_{C\subset \Gamma} s_C \prod_{e\in C} t_e, $$
with $s_C =(\sum_{v\in V(\Gamma_1)} P_v)^2$ and
$P_v=\sum_{e\in E_{ext}(\Gamma), t(e)=v} p_e$, where the momenta satisfy the conservation law  
$\sum_{e\in E_{ext}(\Gamma)} p_e =0$.
The {\em graph polynomial} $\Psi_\Gamma(t)$ is a
homogeneous polynomial of degree $b_1(\Gamma)$ given by
$$ \Psi_\Gamma(t)=\det M_\Gamma(t)=\sum_T \prod_{e\notin T} t_e, $$
with the sum over spanning trees of $\Gamma$, and the matrix 
$$ (M_\Gamma)_{kr}(t)=\sum_{i=0}^n t_i \eta_{ik} \eta_{ir} . $$
Notice how the determinant of this matrix is independent both of the choice of
an orientation of the edges and of a basis of $H_1(\Gamma)$.
Similarly, in the case where $m\neq 0$ but with external momenta $p=0$ one
has
$$ V_\Gamma(t,p)|_{m\neq 0, p=0} = \frac{ m^2 }{\Psi_\Gamma(t)} . $$

After Dimensional Regularization the parametric Feynman integral can be
rewritten as 
$$ U_\mu(\Gamma)(z)=\mu^{-z\ell}\frac{\Gamma(n-\frac{(D+z)\ell}{2})}
{(4\pi)^{\frac{\ell (D+z)}{2}}}  \int_{\sigma_n} \frac{\omega_n}
{\Psi_\Gamma(t)^{\frac{(D+z)}{2}}V_\Gamma(t,p)^{n- \frac{(D+z)\ell}{2}}}  \,\, .
$$ 

Assume for simplicity that we work in the ``stable range" of dimensions $D$
such that $n \leq D\ell/2$, so that we write the integral $U(\Gamma,p)$, up to
a divergent $\Gamma$-factor, in the form
\begin{equation}\label{IntRes}
  \int_{\sigma_n} \frac{P_\Gamma(p,t)^{-n+D\ell/2}}
{\Psi_\Gamma(t)^{-n +(\ell+1)D/2}} \,\omega_n . 
\end{equation}
The integrand is an algebraic differential form on the complement of the
hypersurface
\begin{equation}\label{hatXGamma}
\hat X_\Gamma = \{ t =(t_1,\ldots,t_n) \in \A^n \,|\, \Psi_\Gamma(t)=0 \}.
\end{equation}
Since the polynomial is homogeneous, one can also consider the
projective hypersurface 
\begin{equation}\label{XGamma}
X_\Gamma = \{ t =(t_1:\ldots :t_n) \in \P^{n-1} \,|\, \Psi_\Gamma(t)=0 \}.
\end{equation}
Moreover, the domain of integration is the simplex $\sigma_n$ with
bondary $\partial \sigma_n$ contained in the normal crossings divisor
$\hat\Sigma_n =\{ t \in \A^n \,|\, \prod_i t_i =0 \}$.  Thus, as we discuss
briefly below, if the integral converges, it defines a period of the
hypersurface complement. The integral in general is still divergent, even
if we have already removed a divergent $\Gamma$-factor (hence we are
considering the residue of the Feynman graph $U(\Gamma)$). The
divergences of \eqref{IntRes} come from the intersections
$\hat\Sigma_n \cap \hat X_\Gamma \neq \emptyset$. We discuss later
how one can treat these divergences.

It is worth pointing out here that the varieties $X_\Gamma$ are in general
{\em singular} hypersurfaces, with a singularity locus that is often of low codimension.
This can be seen easily by observing that the varieties defined by the 
derivatives of the graph polynomial
are in turn cones over graph hypersurfaces of smaller graphs and that these
cones do not intersect transversely. Techniques from singularity theory can
be employed to estimate how singular these varieties typically are. Notice how,
from the motivic viewpoint, the fact that they are highly singular is what makes it
possible for many of these varieties (and possibly always for a certain part of
their cohomology), to be sufficiently ``simple" as motives, \ie mixed Tate. This
would certainly not be the case if we were dealing with smooth hypersurfaces.
So the understanding of the singularities of these varieties may play a useful
role in the conjectures on Feynman integrals and motives.

The parametric representation of Feynman integrals and its relation to the
algebraic geometry of the graph hypersurfaces was generalized to theories
with bosonic and fermionic fields in \cite{MaRej} where the analogous result
is obtained in the form of an integration of a Berezinian on a supermanifold.

\subsection{Algebraic varieties and motives} The other main objects involved
in the conjecture on Feynman integrals and periods are {\em motives}. These 
are the focus of a deep chapter of arithmetic algebraic geometry, still in itself
very much at the center of recent investigations in the field. Roughly speaking,
motives are a universal cohomology theory for algebraic varieties, or, to say it
differently, a way to embed the category of varieties into a better (triangulated, 
abelian, Tannakian) category. 

Let $\cV_\K$ denote the category of smooth projective algebraic varieties over 
a field $\K$. For our purposes, we may assume that $\K$ is $\Q$ or a 
number field. The category $\cM_\K$ of pure motives (with the numerical
equivalence relation on algebraic cycles) is defined as
having objects given by triples $(X,p,m)$ of a smooth projective variety $X$,
a projector $p=p^2 \in \End(X)$, and an integer $m\in \Z$. The morphisms 
extend the usual notion of morphism of varieties, by allowing also 
{\em correspondences}, that is, algebraic cycles in the product $X\times Y$.
A morphism in the usual sense is represented by the cycle given by its graph
in $X\times Y$. More precisely, one has
$$ \Hom((X,p,m),(Y,q,n))= q {\rm Corr}^{m-n}_{/\sim}(X,Y) \, p, $$
for projectors $p^2=p$, $q^2=q$, and where ${\rm Corr}^{m-n}(X,Y)$
means the abelian group or vector space of cycles in $X\times Y$ of
codimension equal to $\dim(X)-m+n$ and $\sim$ is the numerical
equivalence relation on cycles (two cycles are the same if they have the
same intersection numbers with any cycle of complementary 
dimension).

One defines the Tate motives $\Q(m)$ by formally setting $\Q(1)=\bL^{-1}$,
the inverse of the Lefschetz motive (the motive of an affine line) and $\Q(m)=\Q(1)^m$,
with $\Q(0)$ the motive of a point, so that $(X,p,m)=(X,p)\otimes \Q(m)$. The
reason for introducing these new objects in the category of motives is to
allow for cycles of varying codimension: this makes it possible to have a
duality $(X,p,m)^\vee=(X,p^t,-m)$ and a rigid tensor structure on the category 
$\cM_\K$. It is known that, with the numerical equivalence on cycles, $\cM_\K$ 
is an abelian category and it is in fact Tannakian. Since it is a semisimple category, 
its Tannakian Galois group (the motivic Galois group) is reductive.
The subcategory generated by the $\Q(m)$ is the category of pure Tate motives,
whose motivic Galois group is $\bG_m$. (See \cite{Andre}, \cite{Jannsen}, \cite{Man-mot}.)

The situation becomes considerably more complicated when the varieties
considered are not smooth projective, for instance, when one wants to
include singular varieties, as is necessarily the case in relation
to quantum field theory, since we have seen that the $X_\Gamma$ are
usually singular varieties. In this case, the theory of motives is not as well
understood as in the pure case. Mixed motives, the theory of motives that
accounts for these more general types of varieties, are known to form a
triangulated category $\cD\cM_\K$, by work of Voevodsky, Levine, Hanamura \cite{Levine},
\cite{Voev}. Distinguished triangles in this triangulated category of motives correspond
to long exact sequences in cohomology of the form
$$ \m(Y) \to \m(X) \to \m(X\smallsetminus Y) \to \m(Y)[1] $$
in the case of closed embeddings $Y\subset X$. Moreover, one has a
homotopy invariance property expressed by the identity
$$ \m(X\times \A^1) =\m (X)(1)[2] . $$

However, in general one does not have an abelian category. The 
subcategory $\cD\cM\cT_\K\subset \cD\cM_\K$ of mixed Tate motives is
the triangulated subcategory generated by the $\Q(m)$. In the case where
$\K$ is a number field, it is known (see \cite{Levine}) that one has a t-structure on $\cD\cM\cT_\K$
whose heart defines an abelian category $\cM\cT_\K$ of mixed Tate motives.
It is in fact a Tannakian category (see \cite{DelGon}), whose Galois group is of the form
$U\rtimes \bG_m$, where the reductive part $\bG_m$ accounts for the
presence of the pure Tate motives among the mixed ones, while $U$ is
a pro-unipotent affine group scheme which accounts for the nontrivial
extensions between pure Tate motives. 

More concretely, examples of mixed Tate motives are given for instance
by algebraic varieties that admit a stratification where all the strata are
built out of locally trivial fibrations of affine spaces. We will discuss some
explicit examples of this sort below, in the context of quantum field theory.

While explicitly constructing objects in $\cM\cT_\K$ or checking whether
given varieties that define objects in $\cD\cM_\K$ are actually mixed Tate,
\ie whether they give objects in $\cD\cM\cT_\K$ or $\cM\cT_\K$, may in general 
be very difficult, there is an easier way to check the motivic nature of a variety $X$
by looking at its class in the Grothendieck ring of varieties $K_0(\cV_\K)$. This
is generated by isomorphism classes $[X]$, subject to the inclusion-exclusion
relation $[X] = [Y] + [X\smallsetminus Y]$ for closed embeddings $Y\subset X$
and with the product given by $[X][Y]=[X \times Y]$. 

The class in the Grothendieck ring can be thought of as a {\em universal 
Euler characteristic} for algebraic varieties, \cite{Bitt}. In fact, additive invariants
of varieties, \ie invariants with values in a commutative ring $R$ which
satisfy $\chi(X)=\chi(Y)$ if $X\cong Y$ are isomorphic varieties,
$\chi(X)=\chi(Y)+\chi(X\smallsetminus Y)$, for closed embeddings 
$Y\subset X$, and are compatible with products, $\chi(X\times Y)=\chi(X)\chi(Y)$,
correspond to ring homomorphisms $\chi: K_0(\cV)\to R$. Examples of
additive invariants are the usual Euler characteristic, or the motivic Euler
characteristic of Gillet--Soul\'e \cite{GilSou}, $\chi: K_0(\cV_\K) \to K_0(\cM_\K)$ with
values in the Grothendieck ring of the category of motives, defined
on projective varieties by $\chi(X)=[(X,id,0)]$ and on more general
varieties in terms of a complex in the category of complexes over $\cM_\K$.

If one denotes by $\bL=[\A^1] \in K_0(\cV_\K)$ the Lefschetz motive,
then the part of $K_0(\cV_\K)$ generated by the Tate motives is a polynomial
ring $\Z[\bL]$ (or $\Z[\bL,\bL^{-1}]$ after formally inverting the Lefschetz motive in
$K_0(\cM_\K)$).
Checking that the class $[X]$ of a variety $X$ lies in this subring gives 
strong evidence for $X$ being a mixed Tate motive. It may seem 
that a lot of information is lost in passing from objects in $\cD\cM_\K$
to classes in $K_0(\cV_\K)$, since in this ring does not retain the information on the
extensions but only keeps the rough information on scissor relations. However,
at least modulo standard conjectures on motives, knowing that the class $[X]$ lies
in the Tate subring $\Z[\bL,\bL^{-1}]$ of $K_0(\cM_\K)$
should in fact suffice to know that the motive is mixed Tate. In any case,
computing in $K_0(\cV_\K)$ provides a lot of useful information on the
motivic nature of given varieties. 

One last thing that we need to recall briefly is the notion of period, as in \cite{KoZa}.
A period is a complex number that can be obtained by pairing via integration
$$ (\omega,\sigma) \mapsto \int_\sigma \omega $$
an algebraic differential form $\omega\in \Omega^{\dim X}(X)$ on an 
algebraic variety $X$ defined over a number field $\K$ with a cycle $\sigma$
defined by semi-algebraic relations (equalities and inequalities) also defined 
over the same field $\K$. If the domain of integration $\sigma$
has boundary $\partial \sigma \neq 0$, then the period should
be thought of as a pairing with a relative homology group
$$ \sigma \in H_{\dim X}(X(\C),\Sigma(\C)), $$
where $\Sigma$ is a divisor in $X$ containing the boundary of $\sigma$.
It is conjectured in \cite{KoZa} that the only relations between periods arise from
the change of variable and Stokes formulae for integrals.

\subsection{The mixed Tate mystery: supporting evidence}

The main conjecture on the relation between quantum fields and motives
can be formulated as follows.

\begin{conj}\label{mixedtateconj}
Are residues of Feynman integrals in scalar field theories always
periods of mixed Tate motives?
\end{conj}

Here ``residues" refers to the removal of the divergent Gamma factor in 
\eqref{UGammaParamFey}. Notice that, in general, the remaining integral still
contains divergences that need to be removed by a renormalization procedure.
Thus, implicit in the above conjecture is also an independence of the regularization
and renormalization scheme used to eliminate divergences.

The supporting evidence for this conjecture starts from 
extensive numerical computations of Feynman integrals
collected by Broadhurst and Kreimer \cite{BroKr}, which showed the
pervasive presence of zeta and multiple zeta values.
This first suggested the fact that mixed Tate motives 
may be involved in this computation, in view of the fact that
multiple zeta values are periods of mixed Tate motives,
according to \cite{Gon2}, \cite{Tera}.

Modulo the serious issue of divergences, the use of Schwinger and
Feynman parameters expresses Feynman integrals as integrations
of an algebraic differential form on the complement of a hypersurface
$X_\Gamma$ in affine space defined by a homogeneous polynomial 
depending on the combinatorics of the graph.

Kontsevich formulated the conjecture that the
graph hypersurfaces $X_\Gamma$ themselves may always be mixed Tate
motives, which would imply Conjecture \ref{mixedtateconj}. 
Although numerically this conjecture was at first verified up to
a large number of loops, Belkale and Brosnan \cite{BeBro} later disproved the conjecture
in general, showing that in fact the $X_\Gamma$ can be arbitrarily
complicated as motives: they proved that the $X_\Gamma$ generate the
Grothendieck ring of varieties. This, however, does not disprove
Conjecture \ref{mixedtateconj}. In fact, even though the varieties
themselves may be more complicated as motives, the part of
the cohomology that is involved in the computation of the 
period may still be a realization of a mixed Tate motive. 

More evidence for the fact that the cohomology involved, that is
the relative cohomology $H^{n-1}(\P^{n-1}\smallsetminus X_\Gamma, \Sigma_n
\smallsetminus (\Sigma_n\cap X_\Gamma))$, where $\Sigma_n$ denotes
the union of the coordinate hyperplanes, is a realization of a mixed Tate
motive was collected by Bloch--Esnault--Kreimer, \cite{BEK}, \cite{Blo}.

More recently, the question has been reformulated by Aluffi--Marcolli \cite{AluMa3}
in terms of a different relative cohomology involving
determinant hypersurfaces and the motives of {\em varieties
of frames}, which gives further evidence for the conjecture, as we explain below.
A different kind of evidence comes from the approach followed in
the work of Connes--Marcolli \cite{CoMa}, where instead of constructing 
motives for specific Feynman graphs, one compares the
``global" properties of the Tannakian category $\cM\cT_\K$ with
a similar category constructed out of the data of perturbative
renormalization, the Tannakian category of {\em flat equisingular
vector bundles}. Although one obtains in this way only a non-canonical
identification between these Tannakian categories, it adds 
evidence to the conjectured relation between perturbative
renormalization and mixed Tate motives.

We give in the following a general overview of these different
methods and results.

\section{A bottom-up approach to Feynman integrals and motives}\label{BupSec}

With these preliminaries in place, we are now ready to discuss more closely 
the two different approaches to the relation of quantum field theory and motives.
We first introduce what we refer to as a ``bottom-up" approach, in the
sense that it deals with the problem on a graph-by-graph basis and
tries, for individual graphs or families of graphs sharing similar
combinatorial properties, to construct explicit associated motives 
and periods computing the Feynman integrals. This approach was
pioneered by the work of Bloch--Esnault--Kreimer \cite{BEK} and
further developed in \cite{Blo}, \cite{BK}. Here I will concentrate
mostly on my recent joint work with Aluffi \cite{AluMa1},
\cite{AluMa2}, \cite{AluMa3}.

As we have mentioned above, the parametric formulation of Feynman
integrals shows that, modulo divergences, they can be written as
periods on the hypersurface complement $\A^n \smallsetminus \hat X_\Gamma$,
with $n=\# E_{int}(\Gamma)$. One can reformulate the integral in the
projective setting. Then the question of whether the period so computed
is a period of a mixed Tate motive can be reformulated as in \cite{BEK}
as the question of whether the relative cohomology
\begin{equation}\label{relcohomXGamma}
H^{n-1}(\P^{n-1}\smallsetminus X_\Gamma,
\Sigma_n\smallsetminus X_\Gamma \cap \Sigma_n)
\end{equation}
is the realization of a mixed Tate motive 
\begin{equation}\label{motiveBEK}
\m(\P^{n-1}\smallsetminus X_\Gamma,
\Sigma_n\smallsetminus X_\Gamma \cap \Sigma_n),
\end{equation}
where $\Sigma_n =\{ t\in \P^{n-1}\,|\, \prod_i t_i =0 \}$ is a
normal crossings divisor containing $\partial \sigma_n$,
the boundary of the domain of integration.

This leads to the question of how complex, in motivic terms,
the graph hypersurfaces $X_\Gamma$ can be. Clearly, if it
were to be the case that these would always be mixed Tate
as motives, then the conjecture on the nature of the period
would follow easily. However, this is known not to be the case,
as we already mentioned above: 
it is known by  \cite{BeBro} that the classes $[X_\Gamma]$ generate the
Grothendieck ring of varieties, hence they cannot all be
contained in the Tate subring $\Z[\bL] \subset K_0(\cV)$. 
The question remains, however, on whether the particular
piece \eqref{relcohomXGamma} may nonetheless be always
mixed Tate even when the variety $X_\Gamma$ itself may
turn out to be more complicated.

One can exhibit explicit examples of computations of
classes $[X_\Gamma]$ in the Grothendieck ring. A useful
method to obtain information on these classes is the
observation, made in \cite{Blo} and used extensively 
in \cite{AluMa1}, \cite{Blo2}, that the classical Cremona
transformation relates the graph hypersurfaces of a
planar graph and its dual graph.

In fact, if $\Gamma$ is a planar graph and $\Gamma^\vee$
denotes the dual graph in a chosen embedding of $\Gamma$, 
the graph polynomials are related by
$$ \Psi_\Gamma(t_1,\ldots, t_n)=(\prod_e t_e) \Psi_{\Gamma^\vee}(t_1^{-1},\ldots, t_n^{-1}). $$
This means that the graph hypersurfaces have the property that
$$ \cC(X_\Gamma \cap (\P^{n-1}\smallsetminus \Sigma_n)) = X_{\Gamma^\vee} \cap
(\P^{n-1}\smallsetminus \Sigma_n), $$
under the Cremona transformation. The latter is defined as
$$ \cC: (t_1:\cdots:t_n) \mapsto (\frac{1}{t_1}:\cdots : \frac{1}{t_n}), $$
which is well defined outside the singularity locus $\cS_n$ of $\Sigma_n$
defined by the ideal $I_{\cS_n}=(t_1\cdots t_{n-1}, t_1\cdots t_{n-2}t_n,\cdots, t_1t_3\cdots t_n)$.
Notice that this relation only gives an isomorphism of the parts of
$X_\Gamma$ and $X_{\Gamma^\vee}$ that lie outside of $\Sigma_n$.

For example, using this method, an explicit formula for the classes $[X_{\Gamma_n}]$
of the hypersurfaces of the infinite family of so called ``banana graphs" were computed
in \cite{AluMa1}. The banana graphs have graph polynomial
$$ \Psi_\Gamma(t) = t_1\cdots t_n (\frac{1}{t_1} + \cdots + \frac{1}{t_n}). $$

\centerline{\fig{0.4}{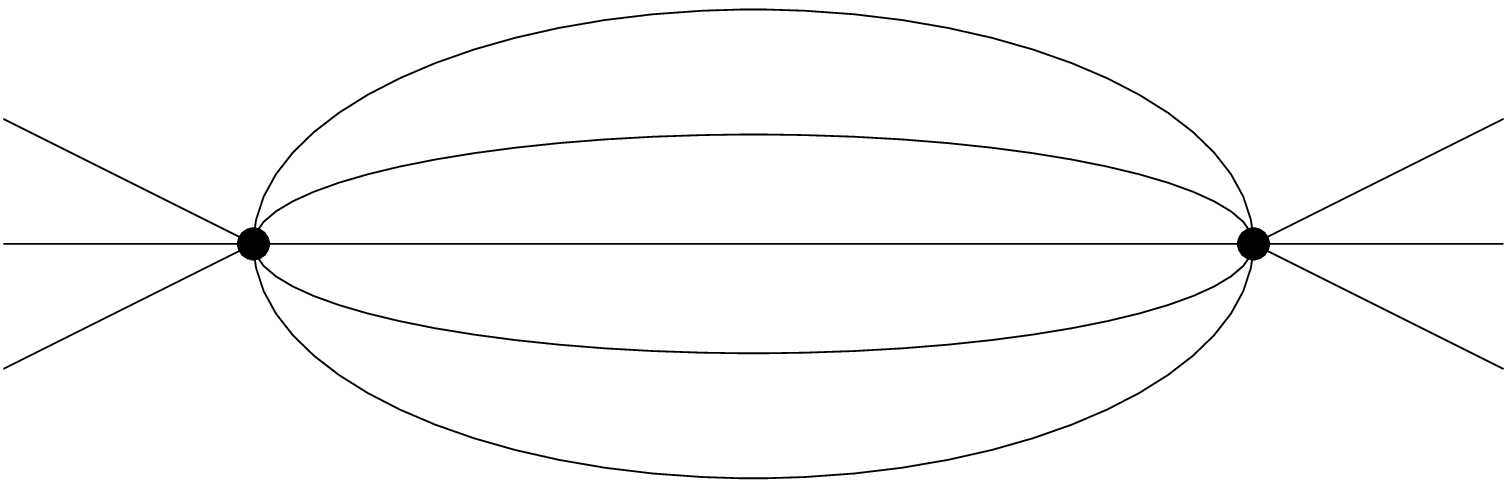}}

The parametric integral in this case is 
$$ \int_{\sigma_n} \frac{(t_1\cdots t_n)^{(\frac{D}{2}-1)(n-1) -1} \,\omega_n}
{\Psi_\Gamma(t)^{(\frac{D}{2}-1)n}} . $$

One has in this case (\cite{AluMa1}) that the class in the Grothendieck ring is
of the form
$$ [X_{\Gamma_n}]= \frac{\bL^n-1}{\bL-1}-\frac{(\bL-1)^n-(-1)^n}{\bL}-n\,
(\bL-1)^{n-2}, $$
so it is manifestly mixed Tate. In fact, in this case the dual graph $\Gamma^\vee$ is
just a polygon, so that $X_{\Gamma^\vee} =\cL$ is a hyperplane in $\P^{n-1}$.
One has
$$ [\cL \smallsetminus \Sigma_n]=[\cL]-[\cL\cap \Sigma_n]= \frac{\bT^{n-1}-(-1)^{n-1}}{\bT+1} $$
where $\bT=[\bG_m]=[\A^1]-[\A^0]$ is the class of the multiplicative group. Moreover, one finds
that  $X_{\Gamma_n}\cap \Sigma_n = \cS_n$ and the scheme of singularities of $\Sigma_n$ has class
$$ [\cS_n] =[\Sigma_n]-n \bT^{n-2} . $$
This then gives
$$ [X_{\Gamma_n}]=[X_{\Gamma_n}\cap \Sigma_n] + [X_{\Gamma_n}\smallsetminus \Sigma_n] ,$$
where one uses the Cremona transformation to identify
$[X_{\Gamma_n}]=[\cS_n]+[\cL\smallsetminus \Sigma_n]$. 

In particular this calculation yields a value for the Euler characteristic of $X_{\Gamma_n}$,
of the form $\chi(X_{\Gamma_n})= n +(-1)^n$. A different computation of the Euler
characteristic based on characteristic classes of singular varieties is also given in \cite{AluMa1}.

A very interesting observation recently made in \cite{Blo2} is that, although individually
the varieties of Feynman graphs may not be mixed Tate, as the result of \cite{BeBro}
shows, cancellations happen when one sums over graphs and one ends up with
a class in $\Z[\bL] \subset K_0(\cV_\K)$. More precisely, it is shown in \cite{Blo2} that
the class
$$ S_N =\sum_{\# V(\Gamma)=N} [X_\Gamma] \frac{N!}{\# \Aut(\Gamma)}  $$
is in $\Z[\bL]$. This is in agreement with the fact that in quantum field theory 
individual Feynman graphs do not represent observable physical processes
and only sums over graphs, usually with fixed external edges and external momenta,
can be physically meaningful. This result suggests that a more appropriate formulation
of the conjecture on Feyman integrals and motives may perhaps be given
directly in terms that involve the full expansion of perturbative quantum field
theory, with sums over graphs, rather than in terms of individual graphs. As
we are going to see below, this also fits in naturally with the other, ``top-down"
approach to relating Feynman integrals to motives that we discuss in the second
half of this paper.

\subsection{Feynman rules in algebraic geometry}

The graph hypersurfaces have another interesting property, namely
the hypersurface complements behave like Feynman rules. This was
first observed and described in detail in the work \cite{AluMa2}, but 
we summarize it here briefly. 

As we recalled above, Feynman rules have certain multiplicative
properties that makes it possible to reduce the combinatorics of
graphs from arbitrary finite graphs to connected and then 1PI graphs,
namely the properties listed in \eqref{Feyrules1} and \eqref{Feyrules2b}.
When working in affine space, one has
$$ \A^{n_1+n_2} \smallsetminus \hat X_\Gamma = (\A^{n_1}\smallsetminus \hat X_{\Gamma_1})
\times (\A^{n_2} \smallsetminus \hat X_{\Gamma_2}), $$ 
for a graph $\Gamma$ that is a disjoint union $\Gamma = \Gamma_1 \amalg \Gamma_2$.
This follows immediately from the fact that the graph polynomial factors as
$$ \Psi_\Gamma(t_1,\ldots, t_n)= \Psi_{\Gamma_1}(t_1,\ldots,t_{n_1}) \Psi_{\Gamma_2}
(t_{n_1+1},\ldots, t_{n_1+n_2}) . $$
In projective space, this would no longer be the case and one has a more
complicated relation in terms of {\em joins} instead of products of varieties,
which gives a fibration
$$ \P^{n_1+n_2-1}\smallsetminus X_\Gamma  \to 
(\P^{n_1-1}\smallsetminus X_{\Gamma_1})\times (\P^{n_2-1}\smallsetminus X_{\Gamma_2}) $$
which is a $\bG_m$-bundle (assuming that $\Gamma_i$ not a forest, else the above map in
projective spaces would not be well defined). Notice that the classes of the affine and the
projective hypersurface complements are related by (\cite{AluMa2})
$$ [\A^n \smallsetminus \hat X_\Gamma] = (\bL-1) [\P^{n-1} \smallsetminus X_\Gamma] , $$
when $\Gamma$ is not a forest, since $[\hat X_\Gamma] = (\bL-1) [X_\Gamma] +1$ is the class
of the affine cone $\hat X_\Gamma$ over $X_\Gamma$.

One can then work either with the Grothendieck ring $K_0(\cV_\K)$ (in which
case one can talk of {\em motivic Feynman rules}), or with a more refined
version where one does not identify varieties up to isomorphisms but only up
to linear coordinate changes coming from embeddings in some ambient 
affine space $\A^N$. This version of Grothendieck ring was introduced in
\cite{AluMa2} under the name of {\em ring of immersed conical varieties} $\cF_\K$.
It is generated by classes $[V]$ of equivalence under linear coordinate changes
of varieties $V\subset \A^N$ (for some arbitrarily large $N$) defined by
homogeneous ideals (hence the name ``conical"), with the usual inclusion-exclusion
and product relations 
$$ [V \cup W] = [V] + [W] - [V\cap W] $$
$$ [V]\cdot [W]= [V\times W] . $$
By imposing equivalence under isomorphisms one falls back on the usual
Grothendieck ring $K_0(\cV)$. The reason for working with $\cF_\K$ instead
is that it allowed us in \cite{AluMa2} to construct invariants of the graph hypersurfaces
that behave like algebro-geometric Feynman rules and that measure to some extent 
how singular these varieties are, and which do not factor through the Grothendieck
ring, since they contain specific information on how the $\hat X_\Gamma$ are embedded
in the ambient affine space $\A^{\# E_{int} (\Gamma)}$.

In general, one defines an $\cR$-valued algebro-geometric Feynman rule, for a
given commutative ring $\cR$, as in \cite{AluMa2} in terms
of a ring homomorphism $I: \cF \to \cR$ by setting
$$  \bU(\Gamma):= I([\A^n])- I([\hat X_\Gamma]) $$
and by taking as value of the {\em inverse propagator} 
$$ \bU(L) = I([\A^1]) . $$
This then satisfies both  \eqref{Feyrules1} and \eqref{Feyrules2b}.
The ring $\cF$ then is the receptacle of the universal algebro-geometric Feynman rule
given by $$\bU(\Gamma)=[\A^n\smallsetminus \hat X_\Gamma]
\in \cF. $$
A Feynman rule defined in this way is {\em motivic} if the homomorphism
$I: \cF \to \cR$ factors through the Grothendieck ring $K_0(\cV_\K)$.

\smallskip

An example of algebro-geometric Feynman rule that does not factor
through $K_0(\cV_\K)$ was constructed in \cite{AluMa2} using the
theory of characteristic classes of singular varieties. 

\smallskip

In the case of smooth varieties, one knows that the Chern classes of the
tangent bundle can be written as a class $c(V)=c(TV)\cap [V]$ in homology 
whose degree of the zero dimensional component satisfies the Poincar\'e--Hopf
theorem $\int c(TV) \cap [V] = \chi (V)$, which gives the topological Euler
characteristic of the smooth variety. This was generalized to singular varieties,
following two different approaches that then turned out to be equivalent, by
Marie--H\'el\`ene Schwartz \cite{Schwartz} and Robert MacPherson \cite{MacPher}. 
The approach followed by
Schwartz generalized the definition of Chern classes as the homology classes of
the loci where a family of $k+1$-vector fields become linearly dependent (for the
lowest degree case one reads the Poincar\'e--Hopf theorem as saying that the
Euler  characteristic measures where a single vector field has zeros). In the
case of singular varieties a generalization is obtained, provided that one assigns
some radial conditions on the vector fields with respect to a stratification with
good properties. The approach of MacPherson was instead based on 
functoriality: a conjecture of Grothendieck--Deligne stated that there should
be a unique natural transformation $c_*$ between the functor $\bF(V)$ of constructible functions on a
variety $V$, whose objects are linear combinations of characteristic classes $1_W$ of
subvarieties $W\subset V$ and where morphisms are defined by the prescription
$f_*(1_W)=\chi(W\cap f^{-1}(p))$, with $\chi$ the Euler characteristic, to the 
homology (or Chow group) functor, which in the smooth case agrees with
$c_*(1_V)=c(TV)\cap [V]$. MacPherson constructed this natural transformation
in terms of data of Mather classes and local Euler obstructions. The results of
Aluffi \cite{Alu} show that, in fact, it is possible to compute these classes without having to
use the original definition and the local data that are usually very difficult to compute.
Most notably, the resulting characteristic classes (denoted $c_{CSM}(X)$ for
Chern--Schwartz--MacPherson) satisfy an inclusion--exclusion formula
$$c_{CSM}(X)=c_{CSM}(Y)+ c_{CSM}(X\smallsetminus Y), $$
but are not invariant under isomorphism, hence they are naturally defined on
classes in $\cF_\K$ but not on $K_0(\cV_\K)$. This classes give a good information
on the singularities of a variety: for example, in the case of hypersurfaces with
isolated singularities, they can be expressed in terms of Milnor numbers, while
more generally for non-isolated singularities, as observed by Aluffi, they can 
be expressed in terms of Euler characteristics of varieties obtained by repeatedly 
taking hyperplane sections. 

To construct a Feynman rule out of these Chern classes, one uses the following
procedure. Given a variety $\hat X\subset \A^N$, one can view it as a locally closed 
locus in $\P^N$, hence one can apply to its characteristic function $1_{\hat X}$
the natural transformation $c_*$ that gives an element in the Chow group $A(\P^N)$ 
or in the homology $H_*(\P^N)$. This gives as a result a class of the form
$$ c_*(1_{\hat X}) = a_0 [\P^0] + a_1 [\P^1] + \cdots + a_N [\P^N] . $$
One then defines an associated polynomial given by (\cite{AluMa2})
$$ G_{\hat X}(T):= a_0 + a_1 T + \cdots + a_N T^N . $$
It is in fact independent of $N$ as it stops in degree equal to $\dim \hat X$.
It is by construction invariant under linear changes of coordinates. It also
satisfies an inclusion-exclusion property coming from the fact that the
classes $c_{CSM}$ satisfy inclusion-exclusion, namely
$$ G_{\hat X \cup \hat Y} (T)=G_{\hat X}(T) + G_{\hat Y}(T) - G_{\hat X \cap \hat Y}(T) $$
It is a more delicate result to show that it is multiplicative, 
$$ G_{\hat X \times \hat Y}(T) = G_{\hat X}(T) \cdot G_{\hat Y}(T). $$
The proof of this fact is obtained in \cite{AluMa2} using an explicit formula
for the CSM classes of joins in projective spaces, where the join $J(X,Y)\subset \P^{m+n-1}$
of two $X\subset \P^{m-1}$ and $Y\subset \P^{n-1}$ is defined as the set of
$$ (sx_1:\cdots: sx_m: ty_1:\cdots: ty_n), \ \ \ \text{ with } \ \  (s:t)\in \P^1 ,$$
and is related to product in affine spaces by the property that the product
$\hat X \times \hat Y$ of the affine cones over $X$ and $Y$ is the affine cone 
over $J(X,Y)$. The resulting multiplicative property of the polynomials
$G_{\hat X}(T)$ shows that one has a {\rm ring homomorphism}
$I_{CSM}: \cF \to \Z[T]$ defined by
$$ I_{CSM}([\hat X]) = G_{\hat X}(T) $$
and an associated Feynman rule 
$$ \bU_{CSM}(\Gamma) = C_\Gamma(T) = I_{CSM}([\A^n]) - I_{CSM}([\hat X_\Gamma]) .$$
This is not motivic, \ie it does not factor through the Grothendieck ring $K_0(\cV_\K)$, as can be
seen by the example given in \cite{AluMa2} of two graphs (see the figure below)
that have different $\bU_{CSM}(\Gamma)$,
$$ C_{\Gamma_1}(T)=T(T+1)^2 \ \ \  C_{\Gamma_2}(T)=T(T^2+T+1) $$
but the same hypersurface complement class in the Grothendieck ring,
$$ [\A^n\smallsetminus \hat X_{\Gamma_i}] =[\A^3]-[\A^2] \in K_0(\cV). $$
\centerline{\fig{0.5}{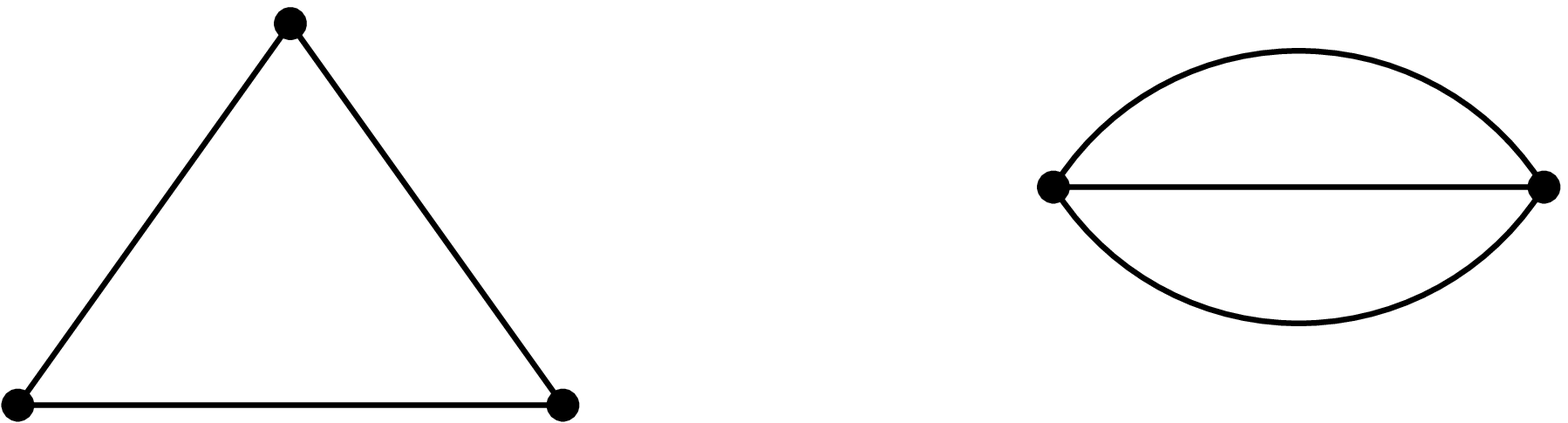}}

\subsection{Determinant hypersurfaces and manifolds of frames}

As our excursion into the algebraic geometry of graph hypersurfaces 
up to this point shows, it seems very difficult to control the complexity 
of the motive 
$$ \m(\P^{n-1}\smallsetminus X_\Gamma,
\Sigma_n\smallsetminus X_\Gamma \cap \Sigma_n) $$
that governs the computation of the parametric Feynman integral as a period.

One way to try to estimate whether the period remains mixed Tate, as the complexity
of the $X_\Gamma$ grows, is to use the properties of periods, in particular the
change of variable formula, which allows one to recast the computation of the
same integral $\int_\sigma \omega$ associated to the data $(X,D,\omega,\sigma)$
of a variety $X$, a divisor $D$, a differential form $\omega$ on $X$, and an
integration domain $\sigma$ with boundary $\partial\sigma \subset D$, 
by mapping it via a morphism $f$ of varieties to another set of data 
$(X',D',\omega',\sigma')$, with the same resulting period whenever 
$\omega = f^*(\omega')$ and $\sigma' = f_*(\sigma)$. In other words, we try to
map the variety $X_\Gamma$ inside a larger ambient variety in such a way that
the part of the cohomology that is involved in the period computation will not
disappear, but the motivic complexity of the new ambient space will be easier 
to control. This is the strategy that we followed in \cite{AluMa3}, which I will
briefly describe here.

The matrix $M_\Gamma(t)$ associated to a Feynman graph $\Gamma$ 
determines a linear map of affine spaces
$$ \Upsilon: \A^n \to \A^{\ell^2}, \ \ \  
\Upsilon(t)_{kr}=\sum_i t_i \eta_{ik} \eta_{ir} $$
such that the affine graph hypersurface is obtained as the preimage
$$ \hat X_\Gamma = \Upsilon^{-1}(\hat \cD_\ell) $$
under this map of the determinant hypersurface 
$$ \hat\cD_\ell =\{ x=(x_{ij}) \in \A^{\ell^2}\,|\, \det(x_{ij})=0 \}. $$
The advantage of moving the period computation via the map $\Upsilon=\Upsilon_\Gamma$
from the hypersurface complement $\A^n \smallsetminus \hat X_\Gamma$
to the complement of the determinant hypersurface $\A^{\ell^2}\smallsetminus \hat\cD_\ell$
is that, unlike what happens with the graph hypersurfaces, it is well known that 
the determinant hypersurface $\hat \cD_\ell$ is a mixed Tate motive.

One can give explicit combinatorial conditions on the graph that ensure 
that the map $\Upsilon$ is an embedding. As shown in \cite{AluMa3}, for
any 3-edge-connected graph with at least 3 vertices and no looping edges, which
admit a closed 2-cell embedding of face width at least 3, the map $\Upsilon$ is injective.
These combinatorial conditions are natural from a physical viewpoint.
In fact, 2-edge-connected is just the usual 1PI condition, while 3-edge-connected
or 2PI is the next strengthening of this condition (the 2PI effective action is
often considered in quantum field theory), and the face width condition
is also the next strengthening of face width 2, which a well known combinatorial
conjecture on graphs \cite{MoTho} expects should simply follow for graphs that are 2-vertex-connected.
(The latter condition is a bit more than 1PI: for graphs with at least two vertices and no looping edges it is 
equivalent to all the splittings of the graph at vertices also being 1PI.) The conditions that
the graph has no looping edges is only a technical device for the proof. In fact, it is then easy
to show (see \cite{AluMa3}) that adding looping edge does not affect the injectivity of the
map $\Upsilon$.

One can then rewrite the Feynman integral (as usual up to a
divergent $\Gamma$-factor) in the form
$$ U(\Gamma) = \int_{\Upsilon(\sigma_n)} \frac{\cP_\Gamma(x,p)^{-n+D\ell/2} \omega_\Gamma(x)}
{\det(x)^{-n+(\ell+1)D/2}} , $$
for a polynomial $\cP_\Gamma(x,p)$ on $\A^{\ell^2}$ that restricts to $P_\Gamma(t,p)$, and
with $\omega_\Gamma(x)$ the image of the volume form. Let then 
$\hat\Sigma_\Gamma$ be a normal crossings divisor in $\A^{\ell^2}$, which
contains the boundary of the domain of integration, 
$\Upsilon(\partial\sigma_n)\subset \hat\Sigma_\Gamma$. The question on the
motivic nature of the resulting period can then be reformulated (again modulo
divergences) in this case as the question of whether the motive
\begin{equation}\label{relmotDet}
 \m (\A^{\ell^2}\smallsetminus \hat\cD_\ell, \hat\Sigma_\Gamma \smallsetminus 
(\hat\Sigma_\Gamma \cap \hat\cD_\ell))  
\end{equation}
is mixed Tate. One sees immediately that, in this reformulation of the question,
the difficulty has been moved from understanding the motivic nature of the
hypersurface complement to having some control on the other term of the
relative cohomology, namely the normal crossings divisor $\hat\Sigma_\Gamma$ and
the way it intersects the determinant hypersurface. One would like to have an argument
showing that the motive of $\hat\Sigma_\Gamma \smallsetminus 
(\hat\Sigma_\Gamma \cap \hat\cD_\ell)$ is always mixed Tate. In that case, knowing
that $\A^{\ell^2}\smallsetminus \hat\cD_\ell$ is always mixed Tate, the
fact that mixed Tate motives form a triangulated subcategory of the triangulated
category of mixed motives would show that the motive \eqref{relmotDet} whose
realization is the relative cohomology would also be mixed Tate.
A first observation in \cite{AluMa3}
is that one can use the same normal crossings divisor $\hat\Sigma_{\ell,g}$ for
all graphs $\Gamma$ with a fixed number of loops and a fixed genus (that is, the
minimal genus of an orientable surface in which the graph can be embedded).
This divisor is given by a union of linear spaces
$$ \hat\Sigma_{\ell,g} = L_1 \cup \cdots \cup L_{\binom{f}{2}} $$ 
defined by a set of equations
$$ \left\{ \begin{array}{rll} x_{ij}& =0 & 1\leq i < j \leq f-1 \\[2mm]
x_{i1}+ \cdots + x_{i,f-1} &  =0 & 1 \leq i \leq f-1 \end{array} \right. $$
where $f=\ell-2g+1$ is the number of faces of an embedding 
of the graph $\Gamma$ on a surface of genus $g$.
A second observation of \cite{AluMa3} is then that, using inclusion-exclusion, 
it suffices to show that arbitrary intersections of the components $L_i$ of 
$\hat\Sigma_{\ell,g}$ have the property that $(\cap_{i\in I} L_i)\smallsetminus \hat\cD_\ell$
is mixed Tate. A sufficient condition is given in \cite{AluMa3} in terms of {\em manifolds
of frames}. These are defined as
$$ \bF(V_1,\ldots,V_\ell) := \{ (v_1,\ldots,v_\ell)\in \A^{\ell^2} \,|\, v_k \in V_k \} $$
for an assigned collection of linear subspaces $V_i$ of a given vector space
$V=\A^{\ell^2}$. If the manifolds of frames are mixed Tate motives for arbitrary
choices of the subspaces, then the desired result would follow. One can check
explicitly the cases of two and three subspaces, for which one has explicit formulae
for the classes $[\bF(V_1,\ldots,V_\ell)]$ in the Grothendieck ring:
$$[\bF(V_1,V_2)]= \bL^{d_1+d_2} - \bL^{d_1} - \bL^{d_2}
-\bL^{d_{12}+1} + \bL^{d_{12}} +\bL ,$$
with $d_i =\dim(V_i)$ and $d_{ij}=\dim(V_i\cap V_j)$, and
$$ [\bF(V_1,V_2,V_3)]= (\bL^{d_1}-1)(\bL^{d_2}-1)(\bL^{d_3}-1) $$
$$ - (\bL-1)((\bL^{d_1}-\bL)(\bL^{d_{23}}-1) + (\bL^{d_2}-\bL)(\bL^{d_{13}}-1) +
(\bL^{d_3}-\bL)(\bL^{d_{12}}-1) $$
$$ + (\bL-1)^2 (\bL^{d_1+d_2+d_3-D} - \bL^{d_{123}+1}) + (\bL-1)^3 $$
which also depends on $d_{ijk}=\dim(V_i\cap V_j \cap V_k)$ and 
$D=D_{ijk}=\dim(V_i+V_j+V_k)$. However, it is difficult to establish an induction
argument that would take care of the cases of more subspaces, and the
combinatorics of the possible subspace arrangements quickly becomes 
difficult to control. 

A reformulation of this problem given in \cite{AluMa3} in terms of 
intersections of unions of Schubert cells in flag varieties suggests a
possible connection to Kazhdan--Lusztig theory  \cite{KaLu}.

\subsection{Handling divergences}

So far we did not discuss how one takes care of the divergences 
caused by the intersections of the graph hypersurface $X_\Gamma$
with the domain of integration $\sigma_n$. The poles of the
integrand that fall inside the integration domain happen necessarily 
along the boundary $\partial \sigma_n$, as in the interior the
graph polynomial $\Psi_\Gamma$ takes strictly positive real values.
Thus, one needs to modify the integrals suitably in such a way as
to eliminate, by a regularization procedure, the intersections
$X_\Gamma \cap \partial\sigma_n$, or (to work in algebro-geometric
terms) the intersections $X_\Gamma \cap \Sigma_n$ which
contains the former. There are different possible ways to
achieve such a regularization procedure. We mention here three
possible approaches. 

One method was developed by Belkale and
Brosnan in \cite{BeBro2} in the logarithmically divergent case where
$n=D\ell/2$, that is, when the polynomial $P_\Gamma(t,p)$ is not
present and only the denominator $\Psi_\Gamma(t)^{D/2}$ appears in
the parametric Feynman integral.  Using Dimensional Regularization,
one can, in this case, rewrite the Feynman integral in the form of a
local Igusa $L$-function 
$$ I(s)= \int_\sigma f(t)^s \omega, $$
for $f = \Psi_\Gamma$. They prove that this $L$-function has a
Laurents series expansion where all the coefficients are periods.
In this setting, the issue of eliminating divergences becomes similar
to the techniques used, for instance, in the context of log canonical 
thresholds. The result was more recently extended to the non-log-divergent case
by Bogner and Weinzierl \cite{BoWei1}, \cite{BoWei2}.

Another method, used in \cite{BEK}, consists of eliminating the divergences 
by separating $\Sigma_n$ and $X_\Gamma$ performing a series of blowups.
Yet another method was proposed in \cite{Mar}, based on deformations instead 
of resolutions. By considering the graph hypersurface $X_\Gamma$ as the
special fiber $X_0$ of a family $X_s$ of varieties defined by the level sets
$f^{-1}(s)$, for $f=\Psi_\Gamma : \A^n \to \A^1$, one can form a tubular neighborhood
$$ D_\epsilon(X)=\cup_{s\in \Delta_\epsilon^*} X_s , $$
for $\Delta_\epsilon^*$ a punctured disk of radius $\epsilon$, 
and a circle bundle $\pi_\epsilon: \partial D_\epsilon(X) \to X_\epsilon$.
One can then regularize the Feynman integral by integrating ``around the
singularities" in the fiber $\pi_\epsilon^{-1}(\sigma \cap X_\epsilon)$. The 
regularized integral has a Laurent series expansion in the parameter $\epsilon$.

In general, as we discuss at length below, a regularization procedure for
Feynman integrals replaces a divergent integral with a function of 
some regularization parameters (such as the complexified dimension
of DimReg, or the deformation parameter $\epsilon$ in the example
here above) in which the resulting function has a Laurent series expansion
around the pole that corresponds to the divergent integral originally considered.
One then uses a procedure of extraction of finite values to eliminate the
polar parts of these Laurent series in a way that is {\em consistent over
graphs}, that is, a renormalization procedure. We therefore turn now to recalling
how renormalization can be formulated geometrically, using the
results of Connes--Kreimer, as this will be the step relating the
``bottom-up" approach to Feynman integrals and motives discussed
so far, to the top down approach developed in \cite{CoMa}, \cite{CoMa2},
\cite{CoMa3}, \cite{CoMa-book}.

\section{The Connes--Kreimer theory}

We give here a very brief overview of the main results of the Connes--Kreimer theory,
as they form the basis upon which the ``top-down" approach to understanding the
relation between quantum field theory and motives rests. As we see more in
detail in the next section, in this context ``top-down" means that the relation between
quantum fields and motives will appear in this second approach from the 
comparison of the formal properties of associated abstract categorical structures 
rather than from a direct comparison of individual objects, as in the approach we
have described in the previous sections.

\subsection{The BPHZ renormalization procedure}

The main steps of what is known in the physics literature as
the Bogolyubov--Parashchuk--Hepp--Zimmermann 
procedure (BPHZ) are summarized as follows. (For more details
the reader is invited to look at Chapter 1 on \cite{CoMa-book}).

\medskip

{\bf Step 1: Preparation:} one replaces the Feynman integral $U(\Gamma)$ of \eqref{zUGammaP}
by the expression
\begin{equation}\label{barR}
\bar R(\Gamma)= U(\Gamma) +\sum_{\gamma \in \cV(\Gamma)} 
C(\gamma) U(\Gamma/\gamma). 
\end{equation}
Here we suppress the dependence on $z$, $\mu$ and the external momenta $p$ for
simplicity of notation. The expression \eqref{barR} is to be understood as a sum of
Laurent series in $z$, depending on the extra parameter $\mu$. The sum is over the
set $\cV(\Gamma)$ all proper subgraphs $\gamma\subset \Gamma$ with the 
property that the quotient graph $\Gamma/\gamma$,
where each component of $\gamma$ is shrunk to a vertex, is still a Feynman graph of the theory.
The main result of BPHZ is that the coefficient of pole of $\bar R(\Gamma)$
is local.

\medskip

{\bf Step 2: Counterterms:} These are the expressions by which the 
Lagrangian needs to be modified to cancel the divergence produced by the graph $\Gamma$.
They are defined as the polar part of the Laurent series $\bar R(\Gamma)$,
$$ C(\Gamma)=-T(\bar R(\Gamma)). $$
Here $T$ denotes the operator of projection onto the polar part of a Laurent series.

\medskip

{\bf Step 3: Renormalized value:} One then extracts a finite value from the integral $U(\Gamma)$
by removing the polar part, not of $U(\Gamma)$ itself but of its preparation:
$$ R(\Gamma) =\bar R(\Gamma) + C(\Gamma) $$
$$ = U(\Gamma) + C(\Gamma) +\sum_{\gamma \in \cV(\Gamma)}  
C(\gamma) U(\Gamma/\gamma) $$

A very nice conceptual understanding of the BPHZ renormalization procedure
with the DimReg+MS regularization was obtained by Connes and Kreimer
\cite{CoKr}, \cite{CoKr2}, based on a reformulation of the BPHZ procedure
in geometric terms.

\subsection{Renormalization, Hopf algebras, Birkhoff factorization}

The first step in the geometric theory of renormalization is the understanding 
that the combinatorics of Feynman graphs of a
given theory is governed by an algebraic structure, which accounts for the bookkeeping 
of the hierarchy of subdivergences that occur in multi-loop Feynman integrals. The
right mathematical structure that describes their interactions is a Hopf algebra.
This was first formulated by Kreimer \cite{Krei1} as a Hopf algebra of rooted trees
decorated by Feynman diagrams, and then by Connes--Kreimer \cite{CoKr}, \cite{CoKr2}
more directly in the form of a Hopf algebra of Feynman diagrams.

The Connes--Kreimer Hopf algebra (\cite{CoKr}) $\cH=\cH(\cT)$ depends on the
choice of the physical theory, in the sense that it involves only graphs that are
Feynman graphs for the specified Lagrangian $\cL(\phi)$. As an algebra it is the
free commutative algebra with generators the 1PI Feynman 
graphs $\Gamma$ of the theory. It is graded, by loop number, or by the number
of internal lines,
$$ \deg(\Gamma_1\cdots\Gamma_n)=\sum_i \deg(\Gamma_i), 
\ \ \ \deg(1)=0. $$
This grading corresponds to the order in the perturbative expansion. 

The coproduct already reveals a close relation to the BPHZ formulae. It is
given on generators by
$$ \Delta(\Gamma)=\Gamma\otimes 1 + 1 \otimes \Gamma + 
\sum_{\gamma \in \cV(\Gamma)}  \gamma \otimes \Gamma/\gamma, $$
where the sum is over proper subgraphs $\gamma \subset \Gamma$ in
a specific class $\cV(\Gamma)$ determined by the property that the
quotient graph $\Gamma/\gamma$ is still a 1PI Feynman graph of the
theory and that $\gamma$ itself is a disjoint union of 1PI Feynman graphs
of the theory. Unlike $\Gamma$ which is assumed connected, the subgraphs
$\gamma$ can have multiple connected components, in which case the
quotient graph $\Gamma/\gamma$ is the one obtained by shrinking each
component to a single vertex. 

The antipode is defined inductively by
$$ S(X)=-X -\sum S(X') X'' , $$
where $X$ is an element with coproduct
$\Delta(X)=X\otimes 1 + 1 \otimes X + \sum X' \otimes X''$,
where all the $X'$ and $X''$ have lower degrees.

We only recalled how the Connes--Kreimer Hopf algebra is constructed
for scalar field theories. Recently, van Suijlekom showed \cite{WvS}, 
\cite{WvS2}, \cite{WvS3} how to extend it to gauge theories, icorporating
Ward identities as Hopf ideals. 

A commutative Hopf algebra $\cH$ is dual to an {\em affine group scheme} $G$, 
defined by algebra homomorphisms
$$ G(A) = \Hom(\cH,A), $$
for any commutative unital algebra $A$. In the
case of the Connes--Kreimer Hopf algebra this $G$ is called the group of {\em  
diffeographisms} of the physical theory $\cT$ and it was proved in \cite{CoKr} that
it acts by local diffeomorphisms on the coupling constants of the theory.

The complex Lie group $G(\C)$ of complex points of the affine group scheme $G$,
defined as $G(\C)=\Hom(\cH,\C)$, 
is a pro-unipotent Lie group. For such groups, which are dual to graded connected
Hopf algebras that are finite dimensional in each degree, Connes and Kreimer proved
by a recursive formula that is is always possible to have a multiplicative 
Birkhoff factorization 
$$ \gamma(z) = \gamma_-(z)^{-1} \gamma_+(z) $$
of loops $\gamma:\Delta^* \to G$, defined on an infinitesimal disk $\Delta^*$
around the origin in $\C^*$, in terms of two holomorphic functions $\gamma_\pm(z)$
respectively defined on $\Delta$ and on $\P^1(\C)\smallsetminus \{ 0 \}$.
The factorization is unique upon fixing a normalization condition $\gamma_-(\infty)=1$.
Notice that such Birkhoff factorizations do not always exist for other kinds of
complex Lie groups, as one can see in the example of $\GL_n(\C)$ where the
existence of holomorphic vector bundles on the Riemann sphere is an obstruction.

In Hopf algebra terms, one can describe a loop $\gamma : \Delta^* \to G(\C)$
on an infinitesimal punctured disk $\Delta^*$ as an algebra homomorphism
$\phi \in \Hom(\cH, \C(\{ z \}))$ with values in the field of germs of meromorphic 
functions (covergent Laurent series). The two terms $\gamma_+$ and
$\gamma_-$ of the Birkhoff factorization are, respectively, algebra homomorphisms
$\phi_+\in \Hom(\cH,\C\{ z\})$ to convergent power series, and $\phi_-\in
\Hom(\cH,\C[z^{-1}])$.
The BPHZ recursive formula is then reformulated in \cite{CoKr} \cite{CoKr2} as 
the Birkhoff factorization applied to the loop $\phi(\Gamma) = U(\Gamma)$
given by the dimensionally regularized unrenormalized Feynman integrals.
In fact, the recursive formula of Connes and Kreimer for the Birkhoff
factorization can be written as
$$ \phi_-(X) = -T( \phi(X) + \sum \phi_-(X') \phi(X'') ), $$
for $\Delta(X) = X\otimes 1 + 1\otimes X + \sum X'\otimes X''$, and with $T$ the
projection onto the polar part of the Laurent series, and
$\phi_+ (X) =\phi(X) + \phi_-(X)$.
The fact that the $\phi_{\pm}$ obtained in this way are still algebra
homomorphism depends on the fact that the projection onto the polar
part of Laurent series is a Rota--Baxter operator. In fact, this renormalization
procedure by Birkhoff factorization was easily generalized in \cite{EFGK}, 
\cite{EFGBP} to arbitrary algebra
homomorphisms $\phi \in \Hom(\cH,\cA)$ from a commutative graded
connected Hopf algebra to a Rota--Baxter algebra.
When one applies this formula to $\phi(\Gamma)=U(\Gamma)$ one
finds the BPHZ formula with $\phi_-(\Gamma)=C(\Gamma)$ the
counterterms and $\phi_+(\Gamma)|_{z=0} = R(\Gamma)$ the
renormalized values.

Notice how, from this point of view, the algebro-geometric Feynman
rules discussed above, correspond to the data of a Hopf algebra
homomorphism $\phi\in \Hom(\cH,\cF_\K)$ or, in the motivic case,
$\phi\in \Hom(\cH, K_0(\cV_\K))$,
together with the assignment of the propagator $\bU(L)=\bL$.
It would therefore be interesting to know if the rings $\cF_\K$ and $K_0(\cV_\K)$
have a non-trivial Rota--Baxter structure.

\section{A top-down approach via Galois theory}\label{TdownSec}

As we mentioned earlier, the ``top-down" approach to the question of
Feynman integrals and periods of mixed Tate motives consists of
comparing categorical structures, instead of looking at 
varieties and motives associated to individual Feynman graphs.
The main idea, developed in my joint work with Connes in 
\cite{CoMa}, \cite{CoMa2}, \cite{CoMa3}, \cite{CoMa-book}, is
to show that the data of perturbative renormalization can be 
reformulated in terms of a Tannakian category of equivalence
classes of differential systems with irregular singularities.

A neutral Tannakian category $\cC$ 
is an abelian category, which is $k$-linear for some
field $k$, has a rigid tensor structure and a fiber functor
$\omega: \cC \to Vect_k$, which is a faithful exact tensor
functor to the category of vector spaces over the same 
field $k$.

Tannakian categories are extremely rigid structures, namely
such a category is equivalent to a category of finite dimensional
linear representations of an affine group scheme, 
$$ \cC \simeq {\rm Rep}_G . $$
The affine group scheme $G$ is
reconstructed from the category as the invertible natural transformations
of the fiber functor.

Thus, in order to relate two sets of objects of a seemingly very different nature, 
of which one is known (as is the case for mixed Tate motives over
a number field) to form a Tannakian category, it suffices to
show that the other set of objects can also be organized in a
similar way, and check that the resulting affine group schemes
are isomorphic: this gives then an equivalence of categories.
This is precisely what is done in the results of \cite{CoMa}.

The reason why this does not yet give an answer to
the conjecture lies in the fact that one only obtains in
this way a non-canonical identification, which cannot
therefore be used to explicitly match Feynman integrals
to mixed Tate motives. There are other mysterious 
aspects, for instance the category of mixed Tate
motives involved in the result of \cite{CoMa} is
not over $\Q$ or $\Z$, but over the ring $\Z[i][1/2]$, while
all the varieties $X_\Gamma$ involved in the parametric 
formulation of Feynman integrals are defined
over $\Z$. Relating explicitly the top-down approach 
described below to the bottom-up approach 
is still an important missing ingredient in the 
geometric theory of renormalization, which may
possibly provide the key to completing a proof
of the main conjecture.

The main results of \cite{CoMa}, \cite{CoMa2}, \cite{CoMa3}
are summarized as follows.
\begin{itemize}
\item Step 1: {\em Counterterms as iterated integrals.} 
One writes the negative piece $\gamma_-(z)$ of the Birkhoff factorization
as an iterated integral depending on a single element $\beta$
in the Lie algebra $Lie(G)$ of the affine group scheme dual to
the Connes--Kreimer Hopf algebra. This is a way of formulating 
what is known in physics as the 't Hooft--Gross relations \cite{Gross}, that is,
the fact that counterterms only depend on the beta function of
the theory (the infinitesimal generator of the renormalization group flow).
\medskip
\item Step 2: {\em From iterated integrals to solutions of irregular singular differential equations.}
The iterated integrals obtained in the first step are uniquely solutions to certain
differential equations. This makes it possible to classify the divergences of
quantum field theories in terms of families of differential systems with singularities.
The fact that, by dimensional analysis, counterterms are independent of the energy 
scale corresponds in these geometric terms to the flat singular connections describing
the differential systems satisfying a certain {\em equisingularity} condition.
\medskip
\item Step 3: {\em Equisingular vector bundles.} Instead of working with equisingular
connections in the context of principal $G$-bundles, one can formulate things equivalently
in terms of linear representations and of flat connections on vector bundles. These data
can then be organized in a neutral Tannakian category $\cE$ which is independent of $G$ and
therefore universal for all physical theories. 
\medskip
\item Step 4: {\em The Galois group.} The Tannakian category of flat equisingular
connections is equivalent to a category of representations $\cE\simeq Rep_{\bU^*}$
of an affine groups scheme $\bU^*=\bU\rtimes \bG_m$, where $\bU$ is the
prounipotent affine group scheme dual to the Hopf algebra $\cH_{\bU}=U(\cL)^\vee$,
where $\cL=\cF(e_{-n}; n\in \N)$ is the free graded Lie algebra with one generator 
in each degree. 
\medskip
\item Step 5: {\em Motivic Galois group.} The same group $\bU^*=\bU\rtimes \bG_m$
is known to arise (up to a non-canonical identification) as the motivic Galois group
of the category of mixed Tate motives over the scheme $S=\Spec(\Z[i][1/2])$,
by a result of Deligne--Goncharov \cite{DelGon}.
\end{itemize}

We describe briefly each of these steps below.

\subsection{Counterterms as iterated integrals}

In the Birkhoff factorization, there is in fact a dependence on a mass scale $\mu$,
inherited from the same dependence of the dimensionally regularized Feynman
integrals $U_\mu(\Gamma)$, so that we have
$$ \gamma_\mu(z) =\gamma_-(z)^{-1} \gamma_{\mu,+}(z) , $$
where one knows by reasons of dimensional analysis that
the negative part is independent of $\mu$. This part is written as a
time ordered exponential 
$$ \gamma_-(z)= T e^{-\frac{1}{z} \int_0^\infty \theta_{-t}(\beta) dt}= 
1 + \sum_{n=1}^\infty \frac{d_n(\beta)}{z^n} , $$
where 
$$ d_n(\beta) = \int_{s_1\geq s_2 \geq \cdots \geq s_n \geq 0} \theta_{-s_1}(\beta) \cdots \theta_{-s_n}(\beta) ds_1 \cdots ds_n ,$$
and where  $\beta \in \Lie(G)$ is the beta function, that is,
the infinitesimal generator of renormalization group flow, and
the action $\theta_t$ is induced by the grading of the Hopf algebra by
$$ \theta_u (X)=u^n X, \ \ \ \text{ for } \ \  u\in\bG_m, \ \ \text{ and } \ \ X\in \cH, \ \  \text{ with } \ \ 
 \deg(X)=n ,$$
with generator the grading operator $Y(X)=n X$.
This result follows from the analysis of the renormalization group
in the Connes--Kreimer theory given in \cite{CoKr} \cite{CoKr2}, with the recursive
formula for the coefficients $d_n$ explicitly solved to give the
time ordered exponential above.

The loop $\gamma_\mu(z)$ that collects all the unrenormalized values
$U_\mu(\Gamma)$ of the Feynman integrals satisfies the scaling property
\begin{equation}\label{muscaling}
 \gamma_{e^t \mu}(z)=\theta_{tz} (\gamma_\mu(z)) 
\end{equation}
in addition to the property that its negative part is independent of $\mu$,
\begin{equation}\label{muindep}
 \frac{\partial}{\partial \mu} \gamma_-(z) =0. 
\end{equation}
The Birkhoff factorization is then written in \cite{CoMa} in terms of iterated integrals as
$$ \gamma_{\mu,+}(z)=T e^{-\frac{1}{z} \int_0^{-z\log\mu} 
\theta_{-t}(\beta) dt}\,\, \theta_{z\log\mu} (\gamma_{reg}(z)) .$$
Thus $\gamma_\mu(z)$ is specified by $\beta$ up to an equivalence
given by the regular term $\gamma_{reg}(z)$.
The equivalence corresponds to ``having the same negative part
of the Birkhoff factorization".

\subsection{From iterated integrals to differential systems}

The second step of the argument of \cite{CoMa} goes as follows.
An iterated integral (or time-ordered exponential) 
$g(b)=Te^{\int_a^b \alpha(t)dt}$ is the unique solution of 
a differential equation $dg(t)=g(t)\alpha(t)dt$ with initial
condition $g(a)=1$.  In particular, 
given the differential field $(K=\C(\{z\}),\delta)$ and an affine group scheme $G$,
and the logarithmic derivative
$$ G(K)\ni f \mapsto D(f)=f^{-1} \delta(f) \in \Lie G(K), $$
one can consider differential equations of the form
$D(f)=\omega$, for a flat $\Lie G(\C)$-valued connection 
$\omega$, singular at $z=0\in \Delta^*$. The
existence of solutions is ensured by the condition of 
trivial monodromy on $\Delta^*$
$$ M(\omega)(\ell)=T e^{\int_0^1 \ell^*\omega} =1,  \ \ \ 
\ell \in \pi_1(\Delta^*). $$
These differential systems can be considered up to the
gauge equivalence relation of $D(fh)=Dh + h^{-1} Df \, h$, for
a regular $h\in \C\{z\}$. The gauge equivalence is the same thing
as the requirement considered above that the solutions have the same
negative piece of the Birkhoff factorization,
$$ \omega'=Dh + h^{-1} \omega h \ \  \Leftrightarrow \ \ f^\omega_-=f^{\omega'}_-, $$
where $D(f^\omega)=\omega$ and $D(f^{\omega'})=\omega'$.

\subsection{Flat equisingular connections}

The third step of \cite{CoMa} consists of reformulating the data of the
loops $\gamma_\mu(z)$ up to the equivalence of having the same
negative piece of the Birkhoff factorization in terms of gauge equivalence
classes of differential systems as above. The point here is that one keeps
track of the $\mu$-dependence and of the way $\gamma_\mu(z)$
scales with $\mu$ and the fact that the negative part of the Birkhoff
factorization is independent of $\mu$, as in \eqref{muscaling},
\eqref{muindep}. In geometric terms these conditions are reformulated
in \cite{CoMa} as properties of connections on a principal $G$-bundle
$P=B\times G$ over a fibration $\bG_b \to B \to \Delta$, where $z\in \Delta$
is the complexified dimension of DimReg and the fiber $\mu^z\in \bG_m$ 
over $z$ corresponds to the changing mass scale. The multiplicative
group acts by
$$ u(b,g)=(u(b),u^Y(g)) \ \ \ \forall u\in \bG_m . $$
The two conditions \eqref{muscaling} and \eqref{muindep} correspond to
the properties that the flat connection $\varpi$ on $P^*$ is {\em equisingular},
that is, it satisfies:
\begin{itemize}
\item Under the action of $u\in \bG_m$ the connection transforms like
$$ \varpi(z,u(v))=u^Y (\varpi (z,u)) . $$
\item If $\gamma$ is a solution in $G(\C(\{z\}))$ of the equation
$D\gamma =\varpi$, then the restrictions
along different sections $\sigma_1,\sigma_2$ of $B$ with
$\sigma_1(0)=\sigma_2(0)$ have ``the same type of singularities", 
namely
$$ \sigma_1^*(\gamma)\sim \sigma_2^*(\gamma) ,$$
where $f_1\sim f_2$ means that $f_1^{-1} f_2\in G(\C\{z \})$, regular at zero.
\end{itemize}

\subsection{Flat equisingular vector bundles} 

The fourth step of \cite{CoMa} consists of transforming the information
obtained above from equivalence classes of flat equisingular connections
on the principal $G$-bundle $P$ to a category $\cE$ of flat equisingular 
vector bundles. This is possible without losing any amount of information,
since the affine group scheme $G$ dual to the Connes--Kreimer Hopf
algebra of Feynman graphs of a given physical theory is completely determined
by its category $Rep_G$ of finite dimensional linear representations.
Thus, considering all possible flat equisingular vector bundles gives rise
to a category that in particular contains as a subcategory the vector bundles 
that come from finite dimensional representations of $G$, for any $G$ associated to
a particular physical theory, while in itself the category $\cE$ does not depend on
any particular $G$, so it is therefore universal for different physical theories.

The category $\cE$ of flat equisingular vector bundles is defined in
\cite{CoMa} as follows.

The objects $Obj(\cE)$ are pairs $\Theta=(V,[\nabla])$, where
$V$ is a  finite dimensional $\Z$-graded vector space, out of which one
forms a bundle $E=B\times V$. The vector space has a filtration 
$W^{-n}(V)=\oplus_{m\geq n} V_m$ induced by the grading and a
$\bG_m$ action also coming from the grading. The class $[\nabla]$
is an equivalence class of equisingular connections, which are
compatible with the filtration, trivial on the induced graded spaces 
$Gr_{-n}^W(V)$, up to the equivalence relation of $W$-equivalence.
This is defined by $T\circ \nabla_1 =\nabla_2 \circ T$ for some
$T\in \Aut(E)$ which is compatible with filtration and trivial on $Gr_{-n}^W(V)$.
Here the condition that the connections $\nabla$ are equisingular means that 
they are $\bG_m$-invariant and that restrictions of solutions to sections of $B$ 
with the same $\sigma(0)$ are $W$-equivalent. 
The morphisms $\Hom_\cE(\Theta,\Theta')$ are linear maps $T: V\to V'$  
that are compatible with grading, and such that on $E\oplus E'$ the following
connections are $W$-equivalent:
$$ \left(\begin{array}{cc} \nabla' & 0 \\ 0 & \nabla \end{array}\right) 
\stackrel{W-equiv}{\simeq} \left(\begin{array}{cc} \nabla' & T\nabla -\nabla' T \\
0 & \nabla \end{array}\right) . $$

\subsection{The Riemann--Hilbert correspondence} Finally, we proved in \cite{CoMa}
that the category $\cE$ is a Tannakian category 
$$ \cE \simeq Rep_{\bU^*}, \ \ \  \text{ with } \ \ \bU^*=\bU\rtimes \bG_m , $$
where $\bU$ is dual, under the relation $\bU(A)=\Hom(\cH_{\bU},A)$, to the Hopf algebra
$\cH_{\bU}= U(\cL)^\vee$ dual (as Hopf algebra) to the universal enveloping algebra of
the free graded Lie algebra $\cL=\cF(e_{-1},e_{-2},e_{-3},\cdots)$.
The renormalization group ${\bf rg}: \bG_a \to \bU$ is a 1-parameter subgroup with 
generator $e=\sum_{n=1}^\infty e_{-n}$.  In particular, the morphism $\bU \to G$ that
realizes the finite dimensional linear representations of $G$ with equisingular 
connections as a subcategory of $\cE$ is given by mapping the generators 
$e_{-n} \mapsto \beta_n$ to the $n$-th graded piece of the beta function of the
theory, seen as an element $\beta =\sum_n \beta_n$ in the Lie algebra $Lie(G)$.
There are universal counterterms in $\bU^*$
given in terms of a {\em  universal singular frame}
$$ \gamma_\bU(z,v)=T e^{-\frac{1}{z} \int_0^v u^Y(e) \frac{du}{u}} . $$
For $\Theta=(V,[\nabla])$ in $\cE$ there exists unique $\rho\in Rep_{\bU^*}$ such that
$$ D\rho(\gamma_{\bU})\stackrel{W-equiv}{\simeq} \nabla . $$ 

This same affine group scheme $\bU^*$ 
appears in the work of Deligne--Goncharov as the motivic Galois group of 
the category of mixed Tate motives $\cM_S \simeq Rep_{\bU^*}$, with
$S=\Spec(\Z[i][1/2])$, albeit up to a non-canonical identification. This
leads to an identification (non-canonically) of the category $\cE$, which by
the previous steps classifies the data of the counterterms in perturbative 
renormalization, with the category $\cM_S$ of mixed Tate motives.

Cartier conjectured \cite{Cartier} the existence of a Galois group acting on the
coupling constants of the physical theories and related both to the groups of
diffeographisms of the Connes--Kreimer theory and to the symmetries of
multiple zeta values, and he referred to it as a {\em cosmic Galois group}. In this
sense the result of \cite{CoMa} is a positive answer to Cartier's  conjecture,
which identifies his cosmic Galois group with the affine group scheme 
$\bU^*=\bU\rtimes \bG_m$.

\section{The geometry of Dim Reg}

We end this exposition with a brief discussion on the subject of
Dimensional Regularization. In physics this is taken to mean a
{\em formal} extension of the rules of integration of Gaussians 
by setting
$$ \int e^{-\lambda t^2} d^z t := \pi^{z/2} \lambda^{-z/2}, $$
for $z\in \C^*$. This prescription can then be used to make sense
of a larger set of integrations in complexified dimension $z$,
which can be reduced to this Gaussian form by the use of Schwinger
parameters. However, no attempt is made to make sense of an
actual geometry in complexified dimension  $z\in \C^*$. 
We argue here that there are (at least) two possible approaches that can
be used to make sense of spaces in dimension $z$ compatibly with
the prescription for the Gaussian integration. One is based on 
noncommutative geometry and it was proposed first in the unpublished
work \cite{CoMa-anom} and later included in our book \cite{CoMa-book},
while the second approach is based on motives and was proposed
in \cite{Mar}.  The noncommutative geometry approach is based on
the idea of taking a product, in the sense of metric noncommutative 
spaces (spectral triples) of the spacetime manifold over which the
quantum field theory is constructed by a noncommutative space $X_z$
whose dimension spectrum (the most sophisticated notion of dimension
in noncommutative geometry) is given by a single point $z\in \C^*$.
The motivic approach is based also on taking a product, but this
time of the motive associated to an individual Feynman graph
by a projective limit of logarithmic motives ${\rm Log}^\infty$.

In both cases the main idea is to deform the geometry by taking a product
of the original geometry on which the computation of the un-regularized
Feynman integral was performed by a new space, either noncommutative
or motivic, which accounts for the shift of $z$ in dimension. Recently there 
has been a considerable amount of activity in relating noncommutative
geometry and motives (see \cite{CCM} and \cite{CoMa-book}). It would 
be interesting to see if, in this context, there is a way to combine these
two approaches to the geometry of Dimensional Regularization.

\subsection{The noncommutative geometry of DimReg}
The notion of metric space in noncommutative geometry is provided by
{\em spectral triples}. These consist of data of the form
$X=(\cA,\cH,\cD)$, with $\cA$ an associative involutive algebra represented
as an algebra of bounded operators on a Hilbert space $\cH$, together
with a self-adjoint
operator $\cD$ on $\cH$, with compact resolvent, and with the property
that the commutators $[a,\cD]$ are bounded operators on $\cH$, for 
all $a\in \cA$. This structure generalizes the data of a compact Riemannian
spin manifold, with the (commutative) algebra of smooth functions, the
Hilbert space of square integrable spinors and the Dirac operator. It
makes sense, however, for a wide range of examples that are not
ordinary manifolds, such as quantum groups, fractals, noncommutative
tori, etc. For such spectral triples there are various different notions of
dimension. The most sophisticated one is the {\em dimension spectrum}
which is not a single number but a subset of the complex plane
consisting of all poles of the family of zeta functions associated to the
spectral triple, 
$$ {\rm Dim}=\{ s\in \C | \zeta_a(s)=\Tr(a|D|^{-s}) \text{ have poles } \} . $$
These are points where one has a well defined integration theory on
the non-commutative space, the analog of a volume form, given in
terms of a residue for the zeta functions. It is shown in \cite{CoMa-anom}, 
\cite{CoMa-book} that there exists a (type II) spectral triple $X_z$ with the
properties that the dimension spectrum is ${\rm Dim}=\{ z\}$  and that
one recoves the DimReg prescription for the Gaussian integration in
the form
$$ \Tr(e^{-\lambda D_z^2}) = \pi^{z/2} \lambda^{-z/2} .$$
The operator $D_z$ is of the form
$D_z=\rho(z) F |Z|^{1/z}$, where $Z=F|Z|$ is a self-adjoint operator 
affiliated to a type II$_\infty$ von Neumann algebra $\cN$ and
$\rho(z)=\pi^{-1/2} (\Gamma(1+z/2))^{1/z}$, with the spectral measure 
$\Tr(\chi_{[a,b]}(Z))=\frac{1}{2} \int_{[a,b]} dt$, for the type II trace. 
The ordinary spacetime over which the quantum field theory
is constructed can itself be modeled as a (commutative) 
spectral triple
$$ X=(\cA,\cH,\cD) =(\cC^\infty(X),L^2(X,S),\Dirac_X)$$
and one can take a product $X\times X_z$ given by the cup product of
spectral triples (adapted to type II case)
$$ (\cA,\cH,\cD)\cup (\cA_z,\cH_z,D_z) =(\cA\otimes \cA_z,\cH\otimes \cH_z,
\cD\otimes 1 + \gamma \otimes D_z) . $$ 
This agrees with what is usually described in physics as the 
Breitenlohner--Maison prescription to resolve the problem of the
compatibility of the chirality $\gamma_5$ operator with the DimReg
procedure, \cite{BrMa}. The Breitenlohner--Maison prescription consists of
changing the usual Dirac operator to a product, which is indeed of
the form as in the cup product of spectral triples,
$$ \cD\otimes 1 + \gamma \otimes D_z . $$
It is shown in \cite{CoMa-anom} and \cite{CoMa-book}
that an explicit example of a space $X_z$ that can be used to
perform Dimensional Regularization geometrically can be
constructed from the ad\`ele class space, the noncommutative
space underlying the spectral realization of the Riemann zeta
function in noncommutative geometry (see \eg \cite{CCM}),
by taking the crossed product of the partially defined action
$$ \cN=L^\infty(\hat\Z\times\R^*)\rtimes \GL_1(\Q) $$
and the trace
$$ \Tr(f)=\int_{\hat\Z\times \R^*} f(1,a)\, da, $$
with the operator
$$ Z(1,\rho,\lambda)=\lambda, \ \ \  
Z(r,\rho,\lambda)=0, \ r\neq 1 \in \Q^* .  $$

\subsection{The motivic geometry of DimReg} We now explain
briefly the motivic approach to Dimensional Regularization proposed in \cite{Mar}.
The Kummer motives are simple examples of mixed Tate 
motives, given by the extensions 
$$M=[u:\Z\to\bG_m]\in \Ext^1_{\cD\cM(\K)}(\Q(0),\Q(1))$$ 
with $u(1)=q\in \K^*$ and the period matrix
$$ \left(\begin{array}{cc} 1 &  0 \\
\log q & 2\pi i \end{array} \right). $$ 
These can be combined in the form of the Kummer extension of Tate sheaves 
$$\cK\in\Ext^1_{\cD\cM(\bG_m)}(\Q_{\bG_m}(0),\Q_{\bG_m}(1)), $$
$$ \Q_{\bG_m}(1)\to \cK \to \bQ_{\bG_m}(0) \to \Q_{\bG_m}(1)[1] .$$
The {\em logarithmic motives} ${\rm Log}^n={\rm Sym}^n(\cK)$ are defined as
symmetric products of this extension, \cite{Ayoub} \cite{Gon4}. 
They form a projective system and one can take the limit as a pro-motive
$${\rm Log}^\infty=\varprojlim_n {\rm Log}^n. $$
This corresponds to the period matrix 
$$ \left( \begin{array}{cccccc} 1 & 0 & 0 & \cdots & 0 &\cdots \\
 \log(s) & (2\pi i) & 0 & \cdots & 0 & \cdots \\
 \frac{\log^2(s)}{2!} & (2\pi i) \log(s) & (2\pi i)^2 & \cdots & 0 & \cdots\\
\vdots & \vdots & \vdots & \cdots & \vdots & \cdots \\
 \frac{\log^n(s)}{n!} & (2\pi i) \frac{\log^{n-1}(s)}{(n-1)!}
& (2\pi i)^2 \frac{\log^{n-2}(s)}{(n-2)!} & 
\cdots & (2\pi i)^{n-1} & \cdots \\ \vdots & \vdots & \vdots & 
\cdots & \vdots & \cdots \end{array}\right) $$

The graph polynomials $\Psi_\Gamma$ associated to Feynman graphs
define motivic sheaves 
$$ M_\Gamma= (\Psi_\Gamma: \A^n\smallsetminus \hat X_\Gamma \to \bG_m, 
\hat\Sigma_n \smallsetminus (\hat X_\Gamma\cap \hat\Sigma_n), n-1, n-1), $$
viewed as objects  $(f:X\to S,Y,i,w)$ in Arapura's category of motivic sheaves, 
\cite{Arapura}.

Then the procedure of Dimensional Regularization can be see as taking a 
product $M_\Gamma \times {\rm Log}^\infty$ in the Arapura category of the 
motivic sheaf $M_\Gamma$ by the logarithmic pro-motive. The product in
the Arapura category is given by the fibered product
$$ (X_1\times_S X_2 \to S, Y_1\times_S X_2 \cup X_1\times_S Y_2, i_1+i_2,w_1+w_2) .$$
The reason for this identification is that period computations on a fibered products satisfy
$$ \int \pi^*_{X_1}(\omega) \wedge \pi^*_{X_2}(\eta)
= \int  \omega\wedge f_1^* (f_2)_*(\eta) , $$
where the integration takes place on $\sigma_1\times_S \sigma_2$ with 
$\sigma_i \subset X_i$ with boundary $\partial \sigma_i\subset Y_i$, according to
the diagram 
$$ \diagram 
& X_1\times_S X_2 \dlto^{\pi_{X_1}} \drto_{\pi_{X_2}} & \\
X_1 \drto^{f_1} & & X_2 \dlto_{f_2} \\
& S &  \enddiagram $$
This leads to writing the dimensionally regularized parametric Feynman integrals
(at least in the log-divergent case where the term $P_\Gamma(t,p)$ is absent) in
the Igusa $L$-function form $\int_{\sigma} \Psi_\Gamma^z \alpha$ as a period
computation on $M_\Gamma \times {\rm Log}^\infty$.

\subsection*{Acknowledgement}
A large part of what I described in this paper is based on recent joint work 
with Paolo Aluffi and on previous joint work with Alain Connes, whom I thank for
their essential contributions to the development of these ideas and results.
Although parts of this article reflect the content
of the lecture I delivered at the ECM in the summer of 2008, new material
developed in the intervening time was included for completeness and
the actual writing was carried out in the winter 2009, during a stay at MSRI: 
I thank the institute for the hospitality and for support. An informal
discussion with Nikolai Reshetikhin and a talk I gave
in Yongbin Ruan's seminar helped finalizing the form of this exposition. 
The topics covered in this survey will appear in extended form in the
book \cite{Mar2} based on a course given at Caltech in the fall 2008.

\end{document}